\documentclass[aps,showpacs,amsmath,amssymbd,showkeys,nofootinbib,preprint,preprintnumbers]{revtex4}

\usepackage{graphicx}
\usepackage{color}
\usepackage{subfigure}
\usepackage{amsmath}

\def \be  {\begin{equation}}
\def \ee  {\end{equation}}
\def \ee  {\end{equation}}
\def \bea {\begin{eqnarray}}
\def \eea {\end{eqnarray}}

\begin{document}

\preprint{ECTP-2016-06}
\preprint{WLCAPP-2016-06}
\vspace*{3mm}

\title{Strangeness chemical potential from the baryons relative to the kaons particle ratios}

\author{Abdel Nasser Tawfik}
\email{a.tawfik@eng.mti.edu.eg}
\affiliation{Egyptian Center for Theoretical Physics (ECTP), Modern University for Technology and Information (MTI), 11571 Cairo, Egypt}
\affiliation{World Laboratory for Cosmology And Particle Physics (WLCAPP), 11571 Cairo, Egypt}

\author{Magda Abdel Wahab}
\affiliation{Physics Department, Faculty of Women for Arts, Science and Education, Ain Shams University, 11577 Cairo, Egypt}

\author{Hayam Yassin}
\affiliation{Physics Department, Faculty of Women for Arts, Science and Education, Ain Shams University, 11577 Cairo, Egypt}

\author{Eman R. Abo Elyazeed}
\affiliation{Physics Department, Faculty of Women for Arts, Science and Education, Ain Shams University, 11577 Cairo, Egypt}

\author{Hadeer M. Nasr El Din}
\affiliation{Basic Science Department, Modern Academy for Engineering, 11571 Cairo, Egypt}

\begin{abstract}

From a systematic analysis of the energy-dependence of four antibaryon-to-baryon ratios relative to the antikaon-to-kaon ratio, we propose an alternative approach determining the strange-quark chemical potential ($\mu_{\mathrm{s}}$). It is found that $\mu_{\mathrm{s}}$ generically genuinely equals one-fifth the baryon chemical potential ($\mu_{\mathrm{b}}$). An additional quantity depending on $\mu_{\mathrm{b}}$ and the freezeout temperature ($T$) should be added in order to assure averaged strangeness conversation. This quantity gives a genuine estimation for the possible strangeness enhancement with the increase in the collision energy. At the chemical freezeout conditioned to constant entropy density normalized to temperature cubed, various particle ratios calculated at $T$ and $\mu_{\mathrm{b}}$ and the resultant $\mu_{\mathrm{s}}$ excellently agree with the statistical-thermal calculations.

\end{abstract}

\pacs{25.75.Dw, 74.62.-c, 25.75.-q}
\keywords{Particle production in relativistic collisions, Transition temperature variations (phase diagram), Relativistic heavy-ion collisions, Strangeness chemical potential}

\maketitle
\tableofcontents
\makeatletter
\let\toc@pre\relax
\let\toc@post\relax
\makeatother

\section{Introduction}

It is conjectured that the produced particle yields (hadrons) in the high-energy collisions can be well characterized by different statistical-thermal approaches \cite{Tawfik:2014eba}, which - in turn - manifest various properties of the thermal and dense medium of quantum choromodynamics (QCD).  At the state of chemical freezeout, the number of produced particle is assumed to be fixed, while various inelastic scatterings are assumed to cease. In constructing a statistical ensemble of the final state produced particles, different types of chemical potentials should be taken into account. Concretely, the extensive quantity $\mu$ (generic chemical potential) and the intensive quantity $T$ (temperature) are noted to characterize the final state particle production. They are thermodynamical parameters, i.e. cannot be measured, directly \cite{Tawfik:2015kwa}. Their variations with the nucleon-nucleon center-of-mass energy ($\sqrt{s_{\mathrm{NN}}}$) can only be deduced,  phenomenologically, where the measured hadron ratios, for instance, are fitted to the statistical-thermal calculations, in which both quantities (others as well) are taken as free parameters. Accordingly, the baryon chemical potential ($\mu_{\mathrm{B}}$) and the freezeout temperature ($T$) are given as functions of  $\sqrt{s_{\mathrm{NN}}}$ \cite{Tawfik:2015kwa}.  The dependence of $T$ on $\mu_{\mathrm{B}}$ draws the freezeout phase diagram, which is to be characterized by various thermodynamic conditions \cite{Tawfik:2016jzk,FOconditions1,FOconditions2,FOconditions3,Tawfik:2005qn,Tawfik:2004ss,Tawfik:2013eua,Tawfik:2013dba}. In the present calculations, we utilize constant entropy density normalized to the freezeout temperature \cite{Tawfik:2005qn,Tawfik:2004ss}.

Other chemical potentials, such as $\mu_{\mathrm{S}}$, $\mu_{\mathrm{Q}}$ and $\mu_{I_3}$, the strangeness, the electric and the isospin chemical potential, respectively, are estimated from conservations of  corresponding quantum charges.  For example, at given $\mu_{\mathrm{B}}$ and temperature ($T$), $\mu_{\mathrm{S}}$ can be determined under the condition of conserved strange quantum numbers in the heavy-ion collisions, as no strange quarks are present in the colliding hadrons \cite{Tawfik:2004vv}. It has been found that the chemical freezeout $\mu_{\mathrm{S}}$ is almost linearly depending on $\mu_{\mathrm{B}}$ \cite{Tawfik:2014eba,Zhao:2014}. The production rates and phase-space distributions of the strange particles, which are strongly depending on $\mu_{\mathrm{S}}$, are of particular importance. Thus, it is believed to reveal characteristics of the created QCD medium. 

It is conjectured that the antiproton-to-proton ratio in the final state carry essential information about the production of antibaryons and baryons. The antikaon-to-kaon ratio is assumed to reflect the asymmetry between charged mesons and their antiparticles, where $\mu_{\mathrm{b}}$ can be eliminated, while $\mu_{\mathrm{s}}$ becomes dominant. The dependence of antiproton-to-proton on antikaon-to-kaon was reported to play a significant role in observing antihadron-to-hadron asymmetry in central heavy-ion collisions \cite{Song:2013isa}. In the present work, we introduce a systematic analysis of the energy-dependence of four antibaryon-to-baryon ratios normalized to the antikaon-to-kaon ratio measured in different experiments \cite{31,32,36,37,38,39,40,41,42,43,44,45,46,47}. We utilize this in order to propose an alternative approach determining $\mu_{\mathrm{s}}$. It is found that $\mu_{\mathrm{s}}$ is sensitive to $T$, $\mu_{\mathrm{b}}$, and the corresponding antikaon-to-kaon ratios. Both types of chemical potentials, $\mu_{\mathrm{b}}$ and $\mu_{\mathrm{s}}$, count for the quarks chemistry, i.e. the corresponding quarks degrees of freedom are taken into account.

We utilize two different approaches; the hadron resonance gas (HRG) model and the quark combination model (QCM). The earlier is a well-known statistical-thermal approach, where recent compilation of particle data group (PDG) is assumed as the hadronic degrees of freedom (dof). The latter, whose dof are valence quarks, provides a statistical alternative {\it combining} these into confined hadrons. 

In QCM, the strange particles can straightforwardly be determined from the derivative of the grand-canonical partition function with respect to the corresponding chemical potential \cite{Tawfik:2014eba}. We have introduced estimations for the effective strangeness suppression factor ($\lambda_q$) and the ratio between the net quarks from the incident nuclei to the newly produced ones ($\gamma_{\mathrm{net}}$). 

In calculating various particle ratios, the resulting $\mu_{\mathrm{s}}$ shall be utilized. A comparison between this novel approach and the one based on statistical-thermal HRG shall be presented, as well. Having such a generic relation, we hope to solve different puzzles of the particle yields and ratios such as kaon-to-pion horn-like structure and the proton anomaly at high energies.

The present paper is organized as follows. The theoretical approaches shall be presented in section \ref{sec:theor}. This is divided into different topics; the statistical-thermal approaches (section \ref{sec:stm}), the hadron resonance gas model (section \ref{sec:HRG}), and the quark combination model (section \ref{sec:qcm}). Section \ref{sec:results} presents the results. The dependence of antibaryon-to-baryon on antikaon-to-kaon ratios shall be presented in section \ref{sec:prvspr}. The strangeness chemical potential shall be calculated in section \ref{sec:rscp}. The utilization of our approach reproducing various particle ratios shall be outlined in section \ref{sec:rvpr}. Section \ref{sec:pRtYl} compares between various particle ratios calculated at resultant $\mu_{\mathrm{s}}$ and the statistical-thermal calculations, at $\sqrt{s_{\mathrm{NN}}}=200~$GeV. Section \ref{sec:cncl} elaborates the final conclusions.

\section{Theoretical approach}
\label{sec:theor}
 
As discussed in introduction, the appearance of the strangeness dof, e.g. through a biased imbalance in the net number of strange quarks, was proposed as a signature for the new-state-of-matter, the quark-gluon plasma (QGP), because of mass and energy differences taking place in the high-energy collisions, although vanishing net strangeness in the colliding system \cite{SE}. This colored quarks and gluons state, known as QGP, was discovered at RHIC \cite{QGP1} and confirmed at LHC energies \cite{QGP2}. An increasing multiplicity of strange and multi-strange particles such as $K$ ($u \bar{s}$), $\Lambda$ ($u d s$), $\Xi$ ($u s s$) and $\Omega$ ($s s s$), was precisely measured in different experiments.

It was found that the strangeness enhancement increases with increasing the averaged number of participants $\langle N_{\mathrm{part}}\rangle$ \cite{Torrieri:2009mq}. The latter increases with the system size. Furthermore, the particles type and the collision energy have a significant effect, as well \cite{Satz2016}. The statistical-thermal models are well suited for the statistical description of the entire ensemble of produced particles, including strange and multi-strange ones.

In the present calculations, we use $\mu_s$ and $\mu_b$ at the quark level. The switch to their hadronic versions; $\mu_S$ and $\mu_B$, respectively, is straightforward. 

\subsection{Statistical thermal approaches}
\label{sec:stm}
	
At high-energies, the colliding nuclei (hadrons) are likely deconfined into QGP. Such strongly interacting system rapidly expands and cools down. At some critical temperatures, the quarks and gluons combine again forming confined hadrons. This process is known as hadronization, a phase transition from QGP to hadrons. Later on when the number of produced particles is fixed, the system is said freezes out, chemically. 

Different statistical-thermal approaches can be implemented in order to characterize the state of chemical freezeout. The freezeout temperature ($T$) and the baryon chemical potential ($\mu_B$) are well-know thermodynamic parameters to be determined \cite{Tawfik:2012zp}
\begin{eqnarray}
Z(T, \mu, V) &=& \mathtt{Tr} \left[ \exp^{\frac{\mu\, N-H}{T}}\right], \label{eq:lnZ}
\end{eqnarray}
where $H$ is the Hamiltonian of the system, which is given as a summation of the kinetic energies of the relativistic Fermi and Bose particles, counting dof of the confined hadrons and thus implicitly including various types of interactions, at least the ones responsible for the formation of further resonances and so on.

The number density ($n=N/V$) can be deduced from the derivation of Eq. (\ref{eq:lnZ}) with respect to chemical potential of interest
\begin{equation}
n_i (T,\mu_{\mathrm{B}},\mu_{\mathrm{S}},\mu_{\mathrm{Q}})=\frac{g_i}{2 \pi^2} m_i^2\; T\; \lambda_{\mathrm{B}_i}\; \lambda_{\mathrm{S}_i}\; \lambda_{\mathrm{Q}_i}\; K_2\left( \frac{m_i}{T}\right) ,
\label{numdens}
\end{equation}
where $m_i$, $g_i$, $\mathrm{B}_i$, $\mathrm{S}_i$, $\mathrm{Q}_i$ are the $i$-th hadron mass, degeneracy factor, and the baryon, strangeness, and electric charge quantum numbers, respectively. $\lambda_i=\exp(-\mu_i/T)$, with $\mu_i =\mathrm{B}_i \mu_{\mathrm{B}} + \mathrm{S}_i \mu_{\mathrm{S}}+\mathrm{Q}_i \mu_{\mathrm{Q}}$ and $\mu_{\mathrm{B}}$, $\mu_{\mathrm{S}}$, $\mu_{\mathrm{Q}}$ are the baryon, strange and electric charge chemical potential. $K_2$ is the second order modified Bessel function.

The thermodynamical parameters; temperature, chemical potentials of different quantum numbers (baryon, strangeness, electric charge, etc.), and quark occupation factors ($\gamma_q$) can be fixed, phenomenologically. The present work introduces an alternative method for the estimation of $\mu_{\mathrm{s}}$. 

From Eq. (\ref{numdens}), the ratio of anti-particle to particle can straightforwardly be calculated. It is apparent that almost all quantities cancel out with each others except the exponential functions which include chemical potentials normalized to temperature
\bea
&&\frac{n_{\bar{\mathrm{p}}}}{n_\mathrm{p}} = \exp\left(\frac{-2\mu_{\mathrm{B}}}{T}\right), \label{Proton} \\
&&\frac{n_{\bar{\Lambda}}}{n_\Lambda} = \exp\left( \frac{-2(\mu_{\mathrm{B}}-\mu_{\mathrm{S}})}{T}\right), \label{Lambda}\\
&&\frac{n_{\bar{\Xi}}}{n_\Xi} = \exp\left( \frac{-2(\mu_{\mathrm{B}}-\mu_{\mathrm{S}})}{T}\right), \label{Xsi}\\
&&\frac{n_{\bar{\Omega}}}{n_\Omega} = \exp\left( \frac{-2(\mu_{\mathrm{B}}-3\mu_{\mathrm{S}})}{T}\right), \label{Omega}\\
&&\frac{n_{\mathrm{K}^-}}{n_{\mathrm{K}^+}} = \exp\left( \frac{2\mu_{\mathrm{S}}}{T}\right). \label{Kaon}
\eea

From Eqs. (\ref{Proton}) - (\ref{Omega}), a general expression for anti-baryon to baryon ratios can be derived,
\begin{equation}
\frac{n_{\bar{\mathrm{B}}}}{n_{\mathrm{B}}} = \exp\left( \frac{-2(\mu_{\mathrm{B}} -\Delta \mathrm{S} \;\ \mu_{\mathrm{S}})}{T}\right),
\label{antiB_B}
\end{equation}
where $\Delta \mathrm{S}$ stands for net number of the constituents strange quarks in the particle ratios of interest. The factor $2$ stands for the net baryon quantum number. In these expressions, we assume that the antiparticle-to-particle ratios are exclusively depending on the freezeout temperature ($T$) and the baryon chemical potential ($\mu_{\mathrm{B}}$). As discussed, these two parameters can be determined from the statistical fit of various particle ratios measured in different high-energy experiments with the statistical-thermal calculations such as the HRG model, which counts for hadrons dof, section \ref{sec:HRG} \cite{Cleymans:2011pe,Tawfik:2013bza}:
\begin{eqnarray}
T(\mu_{\mathrm{B}})&=& a-b\;\mu_{\mathrm{B}}^2 - c \; \mu_{\mathrm{B}}^4 , \label{T_muB}
\end{eqnarray}
where $a$, $b$, and $c$ are constants; $a=0.166 \pm 0.002$ GeV, $b=0.139\pm 0.016$ GeV$^{-1}$ and $c=0.053\pm 0.021$ GeV$^{-3}$.

From Eq. (\ref{antiB_B}), it is obvious that $\mu_{\mathrm{S}}$ can be related to $\mu_{\mathrm{B}}$ \cite{Zhao:2014},
\begin{equation}
\ln(\mathrm{ratio}) = -2\frac{\mu_{\mathrm{B}}}{T} + \Delta \mathrm{S} \frac{\mu_{\mathrm{S}}}{T}.
\end{equation}
The lhs, $\ln(\mathrm{ratio})$, is to be estimated as functions of $\Delta \mathrm{S}$ for the ${\bar{\Lambda}}/{\Lambda}$, ${\bar{\Xi}}/{\Xi}$, ${\bar{\Omega}}/{\Omega}$ ratios at different energies \cite{Zhao:2014}. We conclude that the strangeness enhancement determines which particle species can be produced and accordingly how the particle ratios depend on $\Delta \mathrm{S}$.

From the quark combination model (QCM), section \ref{sec:qcm}, which relates the number of participant quarks (constituents of participant nucleons out of the colliding hadrons) and the net produced quarks, the kaon ratio can be calculated as functions of the strangeness suppression factors ($\lambda_q$), Eq. (\ref{QCM5}), and the ratio between the net quarks from incident nuclei and the newly produced quarks ($\gamma_{\mathrm{net}}$) \cite{Wang:2013duu}. Accordingly, the baryon ratios shall be determined as functions of the kaon ratio. In the present work, we utilize this in order to derive the strangeness chemical potential at various energies.

\subsection{Hadron resonance gas (HRG) model}
\label{sec:HRG}
 
At the chemical freezeout temperature ($T$), the production of hadrons and resonances in the heavy-ion collisions is conjectured to be controlled by the fireball phase-space and different laws of conservation. The phase space of a produced particle depends on its mass, degeneracy and the available fireball energy and volume. Thus, a grand canonical ensemble can be justified for an ensemble of these produced particles. When treating the hadron resonances as a free gas \cite{Karsch:2003vd,Karsch:2003zq,Redlich:2004gp,Tawfik:2004sw,Tawfik:2004vv}, the hadron resonances are assumed to add to the thermodynamic pressure, especially in the hadronic phase, i.e. below the critical temperature. It has been shown that the thermodynamics of the strongly interacting system can be approximated to an ideal gas composed of recent PDG compilation of all hadrons and resonances with mass up to $2~$GeV \cite{Tawfik:2004sw}
\begin{eqnarray}
\ln Z(T, \mu_i ,V) &=& 
\pm \sum_i \frac{V g_i}{2\pi^2}\int_0^{\infty} k^2 dk \ln\left\{1 \pm \exp\left[\frac{(\mu_i -\varepsilon_i)}{T}\right]\right\},  \label{eq:lnz1}
\end{eqnarray}
where $\varepsilon_i(k)=(k^2+ m_i^2)^{1/2}$ is the $i-$th particle dispersion relation, $g_i$ is the
spin-isospin degeneracy factor and $\pm$ stands for fermions and bosons, respectively.  An excellent description for lattice QCD thermodynamics and various aspects of the particle production was obtained \cite{Tawfik:2014eba}.

The number density can be approximated as, compare with Eq. (\ref{numdens}) \cite{Tawfik:2013bza,Wheaton:2009}
\begin{equation}
n_i(T,\mu)= \frac{g_i}{2 \pi^2} T m_i^2 \lambda_i K_2\left(\frac{m_i}{T}\right)  \exp\left(\frac{- \mu_i}{T}\right). \label{eq:n1}
\end{equation}
The influence of strange quark on the QCD phase diagram and on the chemical freezeout has been analyzed \cite{Tawfik:2004vv}. It was found that the freezeout temperature\footnote{This temperature is assumed to be globally valid for an ensemble of a large number of hadrons and resonances, i.e. this quantity is averaged in such a statistical ensemble. It was shown that the freezeout temperature likely vary with the number of strange quarks included in \cite{Castorina:2014cia,OurBH2}.} decreases with the increase in the strange quark contents, for instance comparing $2$ with $2+1$ lattice QCD calculations confirms this conclusion \cite{Tawfik:2004vv}. But, on the other hand, based on recent lattice QCD simulations \cite{Bellwied:2013cta} and the STAR's beam energy scan \cite{Adamczyk:2017iwn}, this study shall be updated in a future work. The main difference between the particle ratios according to Eq. (\ref{numdens}) and the ones according to Eq. (\ref{eq:n1}) is stemming from the fact that in the HRG approach, Eq. (\ref{eq:n1}), we sum over all available resonances and take into consideration their decays into the particle(s) of interest.

For the sake of completeness, we highlight that some colleagues believe that the HRG results strongly depend on the number of resonances that are taken into account, in particular in the strange sector \cite{1404.6511,1702.01113}. It was pointed out that the resonances reported by PDG might not being enough to count for the effective degrees of freedom below the critical temperature. With {\it missing states}, one refers to hadrons, mostly heavy ones, predicted theoretically. It was proposed to include them as well \cite{1404.6511,1702.01113}. 

From calculations (not shown here), we have found that the missing states aren't contributing considerably to the current calculations, as they are heavy resonances. We think that their inclusion might be relevant to the higher order moments. The present calculations are based on the number density.

\subsection{Quark combination model (QCM)}
\label{sec:qcm}

Opposite to the HRG model, the quark combination model (QCM) is another approach describing the hadronization process out of the quarks dof. We limit the discussion to ratios of the averaged yields of various hadrons (baryons) as a characterized property for the hadronization process \cite{Wang:2013duu}
\begin{eqnarray}
\frac{\langle n_{\bar{\mathrm{p}}}\rangle}{\langle n_{\mathrm{p}}\rangle} &=& \left(\frac{2+\lambda_q}{2+\lambda}\right)^3 \frac{\langle n_{\bar{\mathrm{B}}}\rangle}{\langle n_{\bar{\mathrm{B}}}\rangle}, \label{QCM1a} \\
\frac{\langle n_{\bar{\Lambda}}\rangle}{\langle n_\Lambda\rangle} &=& \left(\frac{2+\lambda_q}{2+\lambda}\right)^3 \left(\frac{\lambda}{\lambda_q}\right) \frac{\langle n_{\bar{\mathrm{B}}}\rangle}{\langle n_{\bar{\mathrm{B}}}\rangle}, \label{QCM2a}\\
\frac{\langle n_{\bar{\Xi}}\rangle}{\langle n_\Xi\rangle} &=& \left(\frac{2+\lambda_q}{2+\lambda}\right)^3 \left(\frac{\lambda}{\lambda_q}\right)^2 \frac{\langle n_{\bar{\mathrm{B}}}\rangle}{\langle n_{\bar{\mathrm{B}}}\rangle}, \label{QCM3a}\\
\frac{\langle n_{\bar{\Omega}}\rangle}{\langle n_\Omega\rangle} &=& \left(\frac{2+\lambda_q}{2+\lambda}\right)^3 \left(\frac{\lambda}{\lambda_q}\right)^3 \frac{\langle n_{\bar{\mathrm{B}}}\rangle}{\langle n_{\bar{\mathrm{B}}}\rangle}.
\label{QCM4a}
\end{eqnarray}
From these expressions, we can deduce a generic expression for any antibaryon-to-baryon ratio
\begin{equation}
\frac{\langle n_{\bar{\mathrm{B}}}\rangle}{\langle n_{\mathrm{B}} \rangle} =\frac{(1-\gamma_{\mathrm{net}})^{3.8}}{(1-\gamma_{\mathrm{net}})^{2.8}+4 \gamma_{\mathrm{net}}},
\label{QCM2b}
\end{equation}
where $\gamma_{\mathrm{net}}$ stands for the so-called net number of quarks, which is determined from the incident nuclei relative to the newly produced ones
\begin{equation}
\gamma_{\mathrm{net}} = \frac{\langle N^{Net}_q \rangle}{\langle N_q \rangle},
\label{QCM3b}
\end{equation}
and $\lambda_q$ is the effective strangeness suppression factor,
\begin{equation}
\lambda = \frac{\lambda_q}{1-\gamma_{\mathrm{net}} \left(1-\frac{\lambda_q}{2}\right)}.
\label{QCM4b}
\end{equation}

We propose following parametrization for $\lambda_q$
\begin{equation}
\lambda_q = (0.0632\pm 0.009) \ln{\sqrt{s_{NN}}} + (0.0664\pm 0.0033), \label{QCM5}
\end{equation}
Accordingly, we can reexpress $\gamma_{\mathrm{net}}$ as
\begin{equation}
\gamma_{\mathrm{net}} = (3.536\pm 0.119) \left(\sqrt{s_{NN}}\right)^{-0.802\pm 0.0147}.
\label{QCM}
\end{equation}
On the other hand, it was found for antikano-to-kaon ratios that \cite{Wang:2013duu}
\begin{equation}
\frac{\langle n_{\mathrm{K}^-}\rangle}{\langle n_{\mathrm{K}^+} \rangle} = (1-\gamma_{\mathrm{net}})\frac{1+0.37 \lambda}{1+0.37\lambda+0.13 \lambda \gamma_{\mathrm{net}}}.
\label{QCM7}
\end{equation}
Equations (\ref{QCM1a})-(\ref{QCM4a}) can be determined.  Accordingly, the particle ratios shall be given in dependence on the ratio of net number of quarks. Both colliding quarks and the newly produced ones are depending on the nucleon-nucleon center-of-mass energy. We find that QCM gives a plausible explanation for the asymmetry between antihadrons and hadrons. Forthermore, the number of quarks seems to play a crucial role.

\section{Results}
\label{sec:results}

\subsection{Baryons-to-Kaons Ratios}
\label{sec:prvspr}

\begin{figure}[!htb]
\includegraphics[width=8.cm]{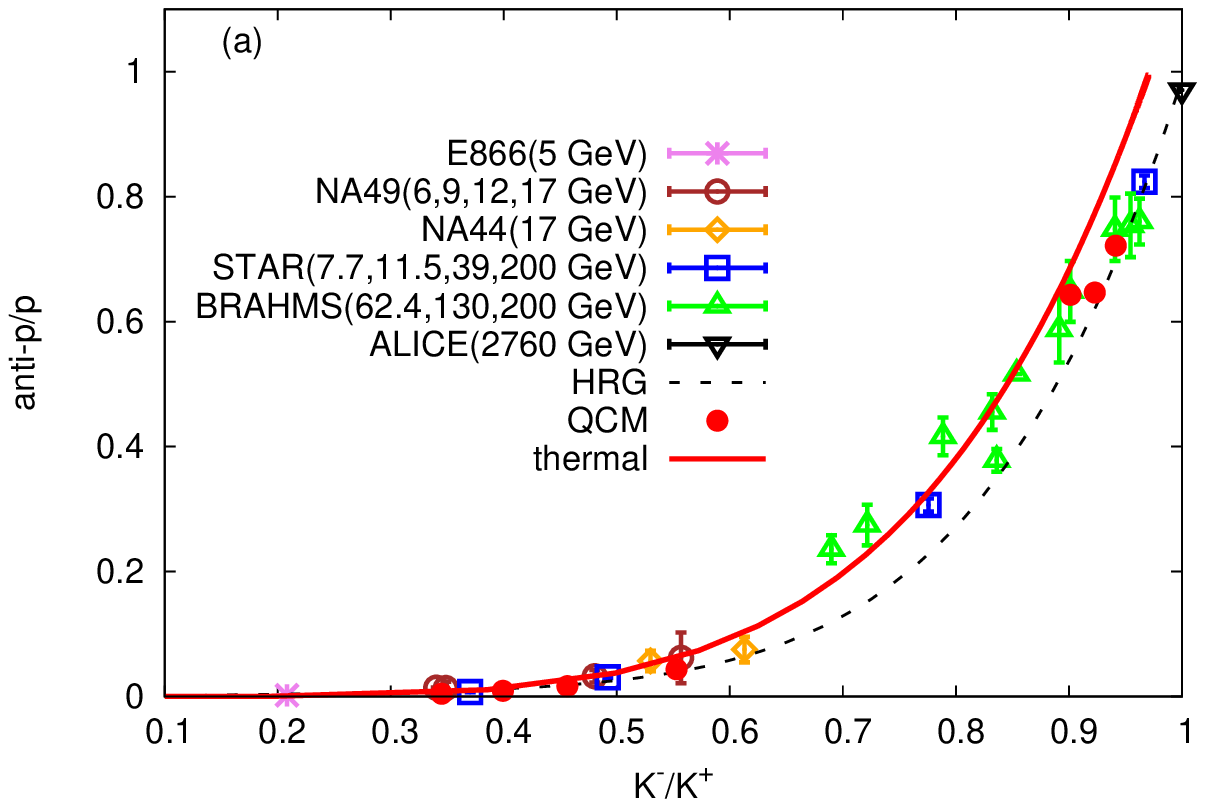}
\includegraphics[width=8.cm]{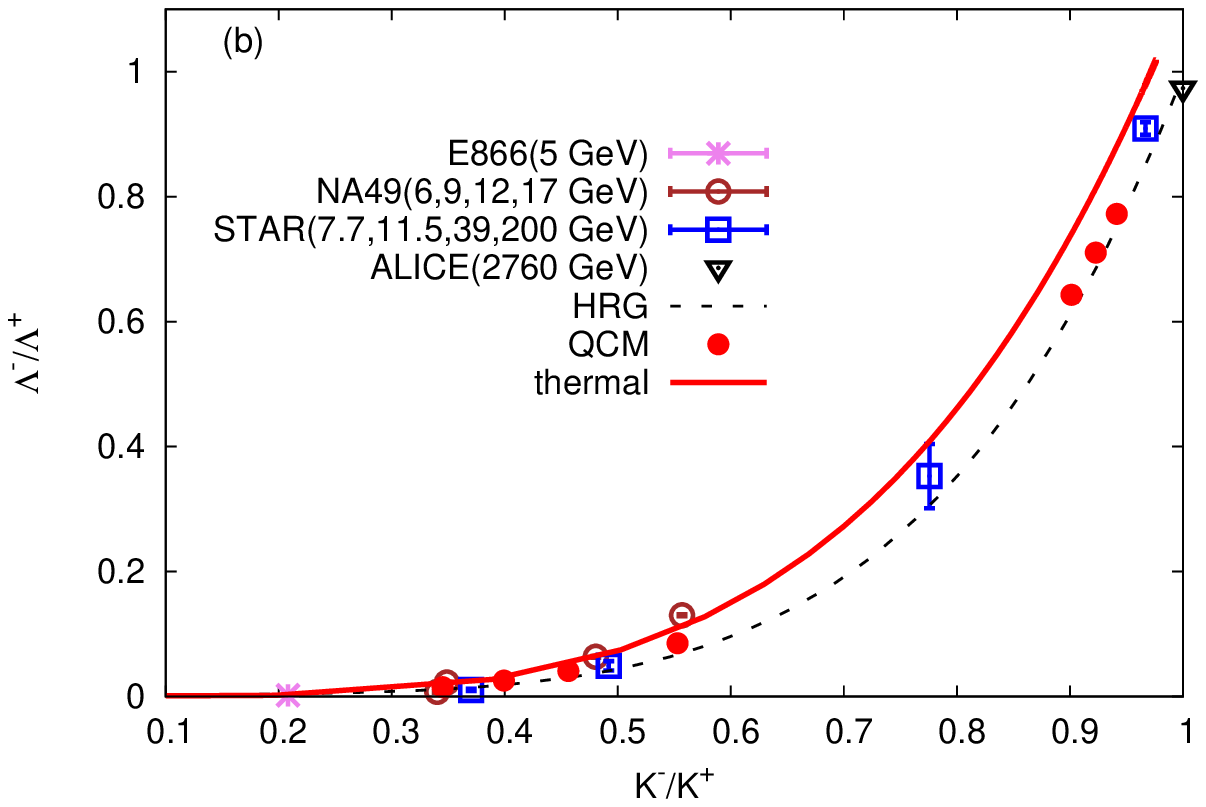}
\includegraphics[width=8.cm]{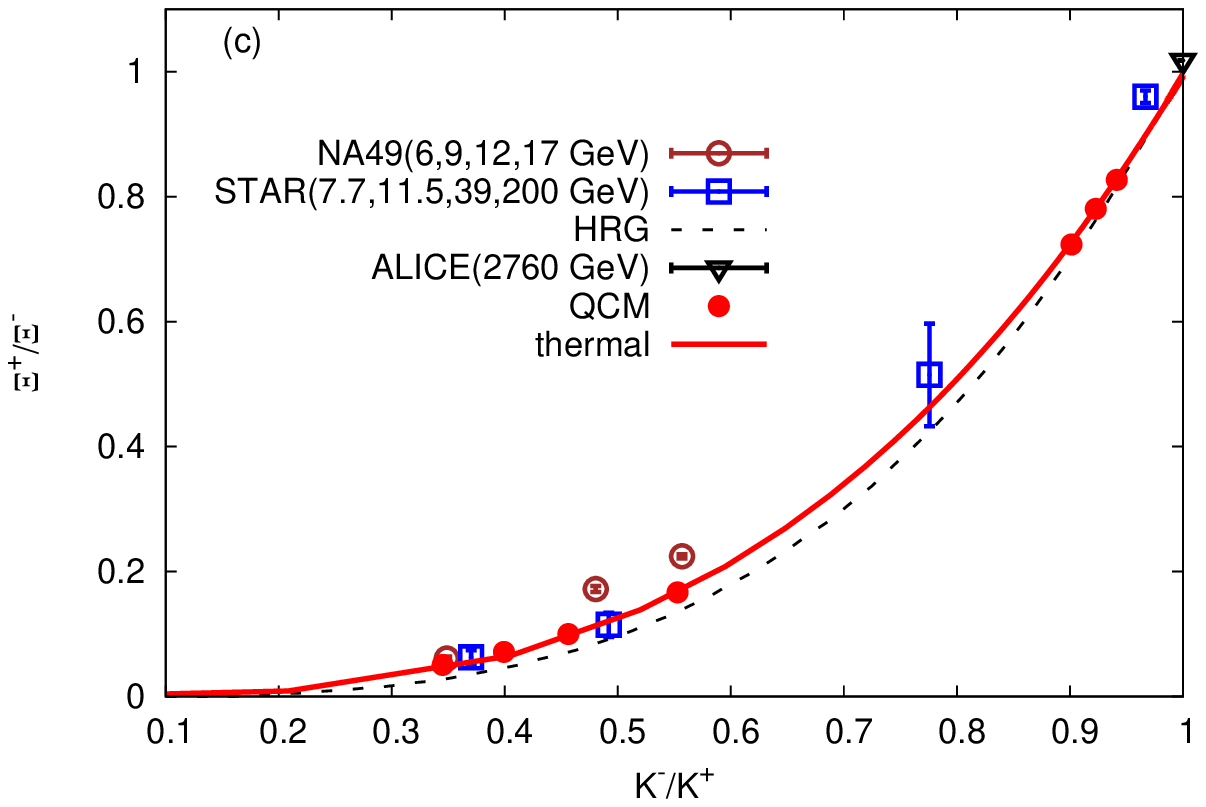}
\includegraphics[width=8.cm]{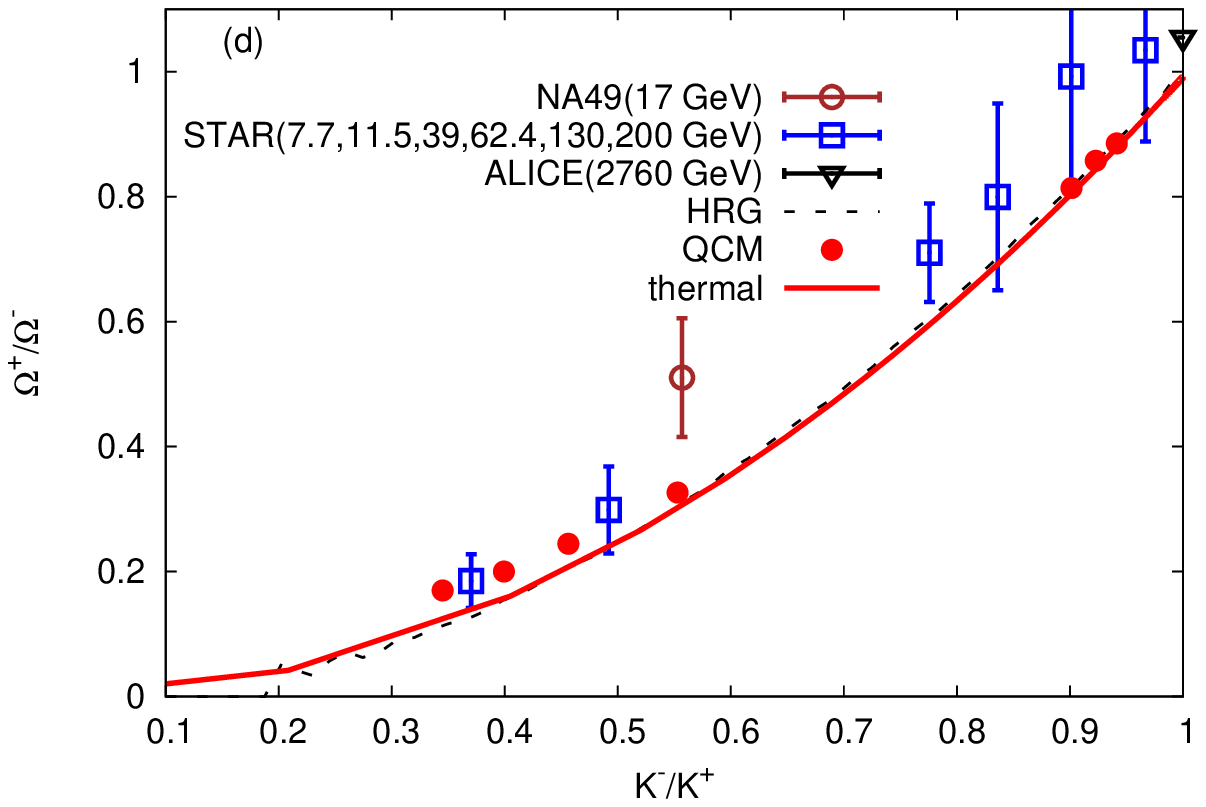}
\caption{The $\mathrm{\bar{B}}/\mathrm{B}$ ratios are given in dependence on $\mathrm{K}^+/\mathrm{K}^-$. The solid curve presents the present calculations, where strange quark $\mu_{\mathrm{s}}$ given by Eq. (\ref{eq:musmubNew2}) should to converted to the hadronic one. The HRG calculations are illustrated by dashed curve, where strange hadron $\mu_{\mathrm{S}}$ is adjusted to assure averaged strangeness conservation. The QCM-calculations are shown as  closed circles. The other symbols refer to the experimental particle ratios; BRAHMS (open triangles), STAR (open squares)  \cite{Bearden:2003fw,Bearden:2001kt,Arsene:2009jg,Kumar:2011us,Abelev:2008ab}, E866 (stars), NA44 (open diamonds), NA49  (open circles) \cite{Ahle:1998xz,Ahle:1999va,Afanasiev:2002mx,Leeuven:2003,Bearden:2002ib} and the ALICE measurements (reciprocal open triangles)  \cite{Abelev:2012wca,Abelev:2013vea}.}
\label{fig:Ratios}
\end{figure}

Figure \ref{fig:Ratios} presents various antibaryon-to-baryon ratios $\mathrm{\bar{B}}/\mathrm{B}$  as functions of the antikaon-to-kaon ratio ($\mathrm{K}^+/\mathrm{K}^-$) at collision energies ranging from $\sqrt{s_{\mathrm{NN}}}=5~$GeV to $2.76~$TeV. The HRG calculations, section \ref{sec:HRG}, are illustrated by the dashed curves. Here, the strange hadron chemical potential $\mu_{\mathrm{S}}$ depends on $T$ and $\mu_{\mathrm{B}}$ so that averaged strangeness conservation is assured. Both chemical potentials directly count for hadrons and resonances. The QCM calculations, section \ref{sec:qcm}, are shown as closed circles. AGS and SPS measurements in the E806 and E866 \cite{Ahle:1998xz,Ahle:1999va}, NA44 \cite{Bearden:2002ib}, and NA49 experiments \cite{Afanasiev:2002mx,Leeuven:2003} are presented as stars, open diamonds, and open circles, respectively. The particle ratios measured at RHIC energies; (BRAHMS) \cite{Bearden:2003fw,Bearden:2001kt,Arsene:2009jg} and (STAR)  \cite{Kumar:2011us,Abelev:2008ab} are given as open triangles and open squares, respectively.  At $\sqrt{s_{\mathrm{NN}}}=2.76~$TeV, the LHC results are illustrated as reciprocal open triangles \cite{Abelev:2012wca,Abelev:2013vea}.

Figure \ref{fig:Ratios}(a) presents $\mathrm{\bar{p}}/\mathrm{p}$ vs. $\mathrm{K}^+/\mathrm{K}^-$ ratios. The RHIC results \cite{Bearden:2003fw,Bearden:2001kt,Arsene:2009jg,Kumar:2011us,Abelev:2008ab} are obtained in different rapidity slices, while the SPS and AGS measurements \cite{Ahle:1998xz,Ahle:1999va,Afanasiev:2002mx,Leeuven:2003,Bearden:2002ib} are performed at midrapidity. LHC measurements for central Pb+Pb collisions at $\sqrt{s_{\mathrm{NN}}}=2.76~$TeV \cite{Abelev:2012wca,Abelev:2013vea} are presented, as well. An excellent agreement is obtained. This combines between all calculations based on QCM (solid circles), and HRG (dashed curve) and the experimental data (symbols). The solid curve represents the present calculations, where $T$ and $\mu_{\mathrm{s}}$ are fixed, phenomenologically, while $\mu_{\mathrm{s}}$ is determined from Eq. (\ref{eq:musmubNew2}). The solid curve depicts our present calculations, where the strange quark $\mu_{\mathrm{s}}$ is so fine-tuned that an  averaged strangeness conservation is guaranteed. $\mu_s$ and $\mu_b$ are straightforwardly converted to $\mu_{\mathrm{S}}$ and $\mu_{\mathrm{B}}$, respectively. 

Figure \ref{fig:Ratios}(b) shows $\bar{\Lambda}/\Lambda$ with respect to $\mathrm{K}^+/\mathrm{K}^-$ ratio, at $\sqrt{s_{\mathrm{NN}}}$ ranging from $5~$GeV to $2.76~$TeV. The RHIC results were measured in the STAR experiment \cite{Zhu:2012ph,Zhao:2014mva}. In central Pb+Pb collisions,  the ALICE measurements at $\sqrt{s_{\mathrm{NN}}}=2.76~$TeV \cite{Milano:2013sza} are also presented. As in panel (a), the HRG-, QCM-calculations are depicted as dashed curve and solid circles, respectively. The solid curve depicts present our calculations for $\bar{\Lambda}/\Lambda$ relative to $\mathrm{K}^+/\mathrm{K}^-$, at varying $\sqrt{s_{\mathrm{NN}}}$, where $\mu_{\mathrm{s}}$ is followed from  Eq. (\ref{eq:musmubNew2})

Figure \ref{fig:Ratios}(c) depicts the various measurements for $\bar{\Xi}/\Xi$ with respect to $\mathrm{K}^+/\mathrm{K}^-$ ratios at collision energies ranging from $\sqrt{s_{\mathrm{NN}}}=6~$GeV to $2.76~$TeV.  RHIC measurements in the STAR experiment \cite{45,32,Zhu:2012ph,Zhao:2014mva} are obtained in different rapidity slices. ALICE measurements in central Pb+Pb collisions at $\sqrt{s_{\mathrm{NN}}}=2.76~$TeV \cite{Milano:2013sza} are also presented. As in panels (a) and (b), HRG, QCM and the statistical-thermal calculations are also depicted. The various curves are similar to the ones described in the top panel.

Figure \ref{fig:Ratios}(d) illustrates $\bar{\Omega}/\Omega$ versus $\mathrm{K}^+/\mathrm{K}^-$ ratios measured at collision energies ranging from $\sqrt{s_{\mathrm{NN}}}=7.7~$GeV to  $2.76~$TeV. The RHIC particle ratios (STAR) \cite{46,47,Abelev:2008ab} are obtained in different rapidity slices, while the SPS results \cite{41,Antinori:2006ij,Bearden:2002ib} are measured at midrapidity. The ALICE particle ratios \cite{Milano:2013sza} are also presented. We observe an excellent agreement combining all calculations (statistical-thermal model, QCM, and HRG model) and the experimental results. The curves are described in the other panels.

\subsection{Strangeness chemical potential obtained}
\label{sec:rscp}

The present work introduces a systematic study for the dependence of the antibaryon-to-baryon ratios ($R_{\mathrm{B}}$) on the antikaon-to-kaon ratios ($R_{\mathrm{K}}$) and therefore proposes a {\it direct} evaluation of the strangeness chemical potential for quarks ($\mu_{\mathrm{s}}$) in a wide range of collision energies. In the statistical-thermal models, it was found that the {\it bulk} freezeout parameters; the temperature ($T$) and the baryon chemical potential of hadrons ($\mu_{\mathrm{B}}$), vary with the number of the strange quarks forming the particle species \cite{Tawfik:2004vv,Castorina:2014cia,OurBH2}. Thus, $\mu_{\mathrm{s}}$ and $\mu_{\mathrm{b}}$ are proposed to count for the hadron constituents.

In Fig. \ref{fig:Ratios}, $\mathrm{\bar{p}}/\mathrm{p}$, $\bar{\Lambda}/\Lambda$, $\bar{\Xi}/\Xi$, and $\bar{\Omega}/\Omega$ are depicted in dependence on $\mathrm{K}^+/\mathrm{K}^-$,  at various $\sqrt{s_{NN}}$ or $\mu_B$ as expressed in Eqs. (\ref{eq:mubSqrts}) and (\ref{mu_B}). 
The QCM calculations (solid circles), section \ref{sec:qcm}, agree well with the experimental results (symbols). The latter are fitted for $\mu_{\mathrm{s}}$ 
\bea
\mathrm{\frac{R_p}{R_K}}:&& \quad  \mu_{\mathrm{s}} = (0.1919\pm 0.007)\, \mu_{\mathrm{b}} + (0.0024\pm 0.0016), \\
\mathrm{\frac{R_{\Lambda}}{R_K}}:&& \quad  \mu_{\mathrm{s}} = (0.1991\pm 0.008)\, \mu_{\mathrm{b}} + (0.0019\pm 0.0016),  \\
\mathrm{\frac{R_{\Xi}}{R_K}}:&& \quad  \mu_{\mathrm{s}} = (0.2294\pm 0.0072)\, \mu_{\mathrm{b}} - (0.001\pm 0.0001), \\
\mathrm{\frac{R_{\Omega}}{R_K}}:&& \quad  \mu_{\mathrm{s}} = (0.2437\pm 0.0028)\, \mu_{\mathrm{b}} - (0.0028 \pm 0.0018).
\eea
The factors of the first terms can be approximately averaged to $0.1991\pm0.01$. A generic expression for the dependence of $\mu_{\mathrm{s}}$ on $\mu_{\mathrm{b}}$ can be proposed
\bea
\mu_{\mathrm{s}} = (0.1991\pm0.01)\, \mu_{\mathrm{b}} + {\cal O}(T(\mu_{\mathrm{b}})), \label{eq:musmubNew}
\eea
where the functionality ${\cal O}(T(\mu_{\mathrm{b}}))$ - in turn - depends on $\sqrt{s_{\mathrm{NN}}}$, as well. The ability of the parametrization to reproduce various particle ratios, for instance, was shown in Fig. \ref{fig:Ratios}.

From the statistical fits of present calculations of the baryons-to-kaons ratios $\mathrm{\bar{p}}/\mathrm{p}$, $\bar{\Lambda}/\Lambda$, $\bar{\Xi}/\Xi$, and $\bar{\Omega}/\Omega$ as functions of $\mathrm{K}^+/\mathrm{K}^-$, with the experimental results at various collision energies, $T$, $\mu_{\mathrm{b}}$, and $\mu_{\mathrm{s}}$ (or equivalently $\mu_{\mathrm{B}}$ and $\mu_{\mathrm{S}}$, respectively) can be determined. The main result is that, $\mu_{\mathrm{s}}$ can be determined from Eq. (\ref{eq:musmubNew}), while the freezeout temperature ($T$) follows Eq. (\ref{T_muB}), and the baryon chemical potential ($\mu_{\mathrm{B}}$) could also be parameterized as,
\bea
\mu_{\mathrm{B}} &=& (4.283\pm 0.245) \left(\sqrt{s_{\mathrm{NN}}}\right)^{-1.025\pm 0.0242}. \label{eq:mubSqrts}
\eea
In this respect, we recall the baryon chemical potential of {\it hadrons} can be related to $\sqrt{s_{\mathrm{NN}}}$ \cite{Andronic2009,Tawfik:2014eba,Tawfik:2013bza}. In GeV units,
\begin{eqnarray}
\mu_{\mathrm{B}} &=& \frac{1.245\pm 0.094}{1+ (0.264\pm 0.028) \sqrt{s_{\mathrm{NN}}}}. \label{mu_B}
\end{eqnarray}

\begin{figure}[!htb]
\includegraphics[width=8.cm]{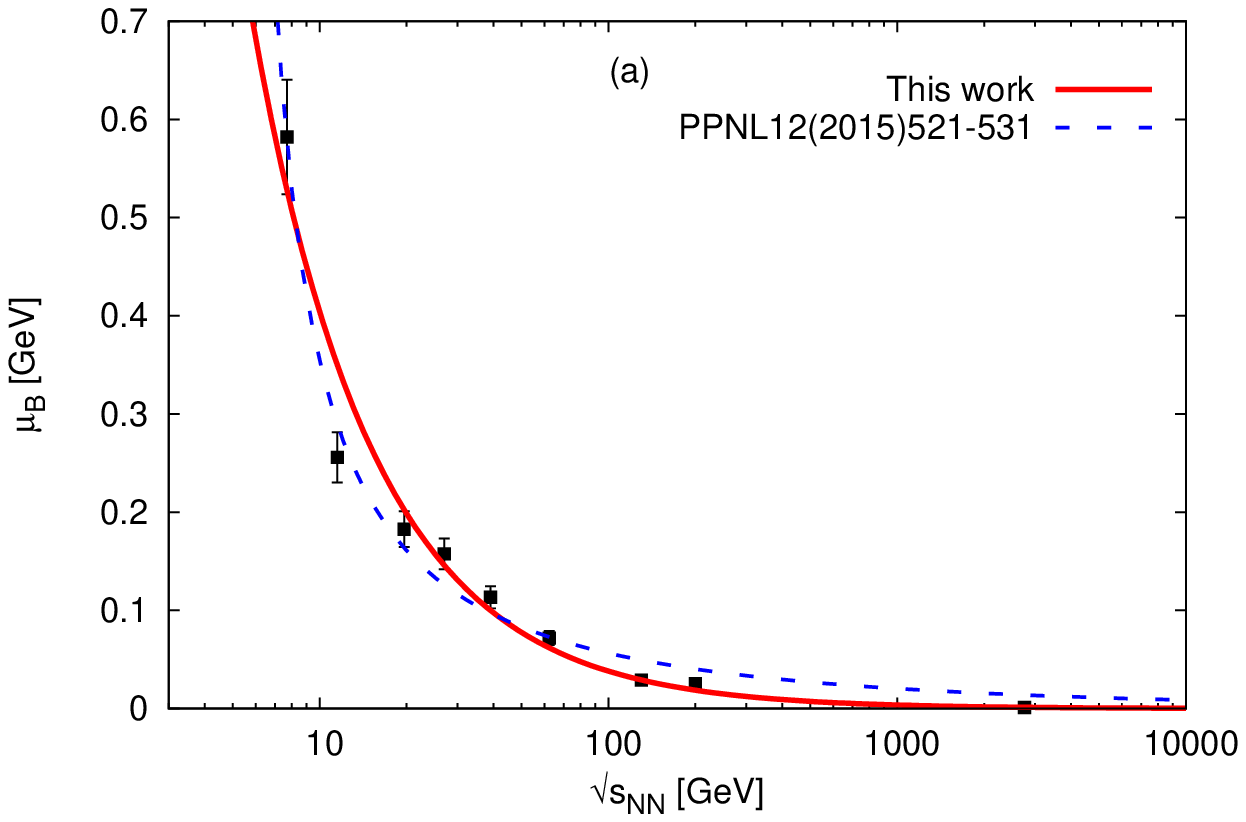}
\includegraphics[width=8.cm]{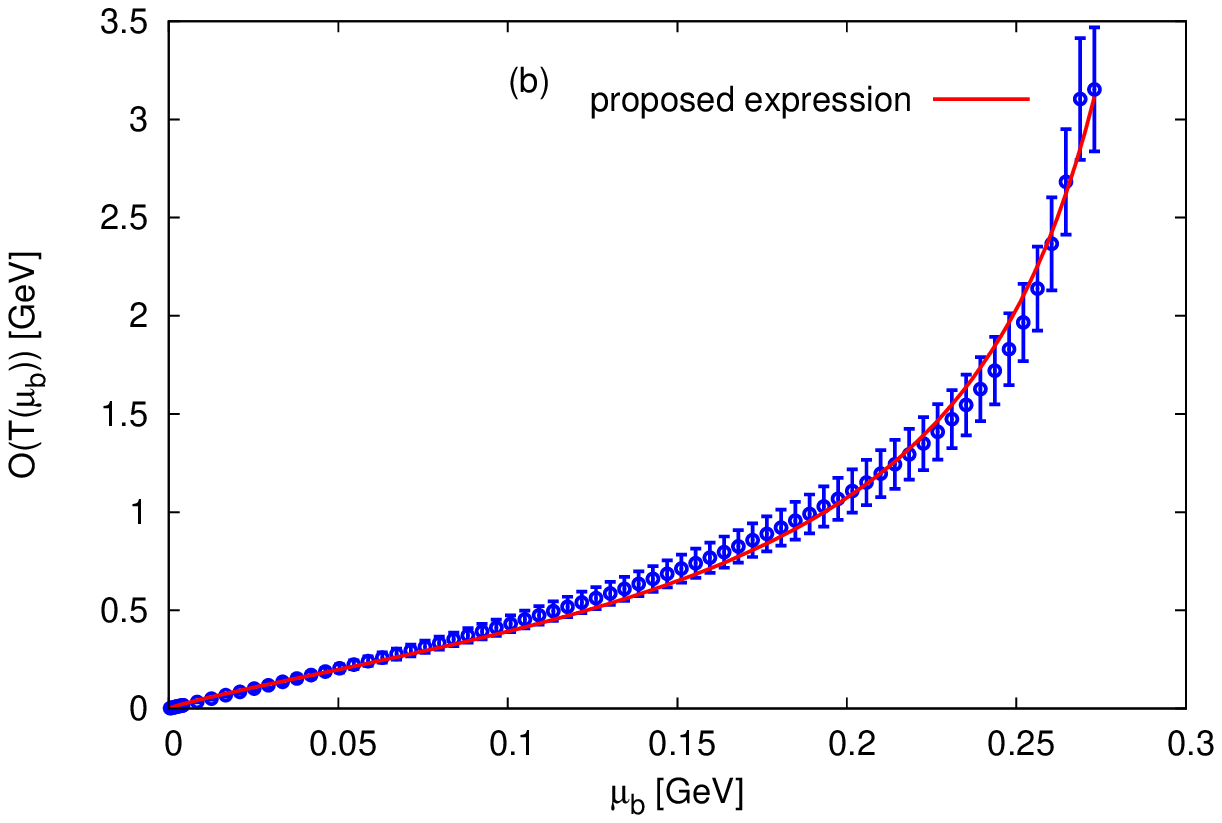} 
\caption{Left-hand panel (a) depicts the dependence of $\mu_{\mathrm{B}}$ on $\sqrt{s_{\mathrm{NN}}}$. The results obtained from the statistical-thermal fits of various baryon-to-kaon ratios (symbols) are well reproduced by Eq. (\ref{eq:mubSqrts}) (solid curve). The dashed line illustrates Eq. (\ref{mu_B})  \cite{Tawfik:2013bza}. Right-hand panel (b) depicts  the energy-dependence of ${\cal O}(T(\mu_{\mathrm{b}}))$, Eq. (\ref{eq:musmubNew}). The symbols with errorbars stand for calculations assuring an overall strangeness conservation. The solid curve represents the parameterization given in the second term of Eq. (\ref{eq:musmubNew2}).}
\label{fig:mubSqrts}
\end{figure}

Left-hand panel (a) of Fig. \ref{fig:mubSqrts} depicts {\it hadronic} $\mu_{\mathrm{B}}$ as functions of $\sqrt{s_{\mathrm{NN}}}$. From the statistical-thermal fits of various baryon-to-kaon ratios (symbols), we obtained partly phenomenological parameterization, Eq. (\ref{eq:mubSqrts}). Interested readers are kindly advised to review Fig. \ref{fig:C} for details on the particle ratios utilized in the present study. The solid curve represents Eq. (\ref{eq:mubSqrts}), while Eq. (\ref{mu_B}) is depicted by the dashed curve \cite{Tawfik:2013bza}. So far, we conclude that both parameterizations, Eqs. (\ref{eq:mubSqrts}) and (\ref{mu_B}), agree well with each other and fairly reproduce the results obtained from the present statistical fit, see Fig. \ref{fig:C}. 

Some details on $\mu_{\mathrm{s}}$ (or equivalently $\mu_{\mathrm{S}}$) utilized in the statistical-thermal models \cite{Tawfik:2014eba} are now in order. At a given temperature ($T$) and baryon chemical potential  ($\mu_{\mathrm{B}}$), $\mu_{\mathrm{S}}$ is so fine-tuned that an averaged strangeness conservation is assured. The resultant $\mu_{\mathrm{S}}$ besides $T$ and $\mu_{\mathrm{B}}$ shall be used in determining various thermodynamic quantities, including particle ratios. Both $T$ and $\mu_{\mathrm{B}}$ and accordingly $\mu_{\mathrm{S}}$ are the free parameters which assure a best agreement (fit) between the calculated particle ratios and the measured ones. Accordingly, $\mu_{\mathrm{S}}$ could be related to $\mu_{\mathrm{B}}$, see for instance \cite{Tawfik:2004sw}. It is obvious that this procedure is pure statistical. No other constrains are restricted. No other ingredients are provided. 

The present work introduces constrains and adds additional ingredients to $\mu_{\mathrm{s}}$, which is first fixed to $\sim20\%$ of $\mu_{\mathrm{b}}$. From Eq. (\ref{eq:musmubNew}), the additional quantity ${\cal O}(T(\mu_{\mathrm{b}}))$ plays the role to regulate the overall dependence of $\mu_{\mathrm{s}}$ on $\mu_{\mathrm{b}}$. The procedure that determines the functionality ${\cal O}(T(\mu_{\mathrm{b}}))$ can be summarized as follows. At a given $T$ and $\mu_{\mathrm{b}}$, $\mu_{\mathrm{s}}$ is first fixed to fifth $\mu_{\mathrm{b}}$. If this doesn't assure the strangeness conservation, ${\cal O}(T(\mu_{\mathrm{b}}))$ comes to play such a role. From statistical-thermal approaches, ${\cal O}(T(\mu_{\mathrm{b}}))$ is determined at various $T$ and $\mu_{\mathrm{B}}$ or $\mu_{\mathrm{b}}$, which can be related to $\sqrt{s_{\mathrm{NN}}}$, Eq. (\ref{eq:mubSqrts}), so that an overall strangeness conservation is guaranteed. This means that the statistical inputs here are limited to ${\cal O}(T(\mu_{\mathrm{b}}))$, while $\mu_{\mathrm{s}}$ is largely fixed to $\sim 0.2\, \mu_{\mathrm{b}}$. 

Right-hand panel of Fig. \ref{fig:mubSqrts} presents ${\cal O}(T(\mu_{\mathrm{b}}))$ vs. $\mu_{\mathrm{b}}$ as calculated from the HRG calculations, where $\sim 0.2\, \mu_{\mathrm{b}}$ is first assigned to $\mu_{\mathrm{s}}$. ${\cal O}(T(\mu_{\mathrm{b}}))$ is varied until an averaged strangeness conservation is assured. The resultant ${\cal O}(T(\mu_{\mathrm{b}}))$ can be parameterized as follows.
\bea
\mu_{\mathrm{s}} = (0.1991\pm0.01) \mu_{\mathrm{b}} + \frac{T(\mu_{\mathrm{B}})}{a_1\, \mu_{\mathrm{b}}^{-b_1} -c_1\, \mu_{\mathrm{b}}}, \label{eq:musmubNew2}
\eea
where $a_1=0.3998\pm 0.006~$GeV$^{-1}$, $b_1=0.856\pm 0.043$, and $c_1=3.278\pm 0.07~$GeV$^{-1}$ and $T(\mu_{\mathrm{B}})$ in GeV units was expressed in Eq. (\ref{T_muB}).

\subsection{Implications}
\label{sec:implc}

\subsubsection{Energy dependence of various particle ratios}
\label{sec:rvpr}

\begin{figure}[!htb]
\includegraphics[width=5.cm]{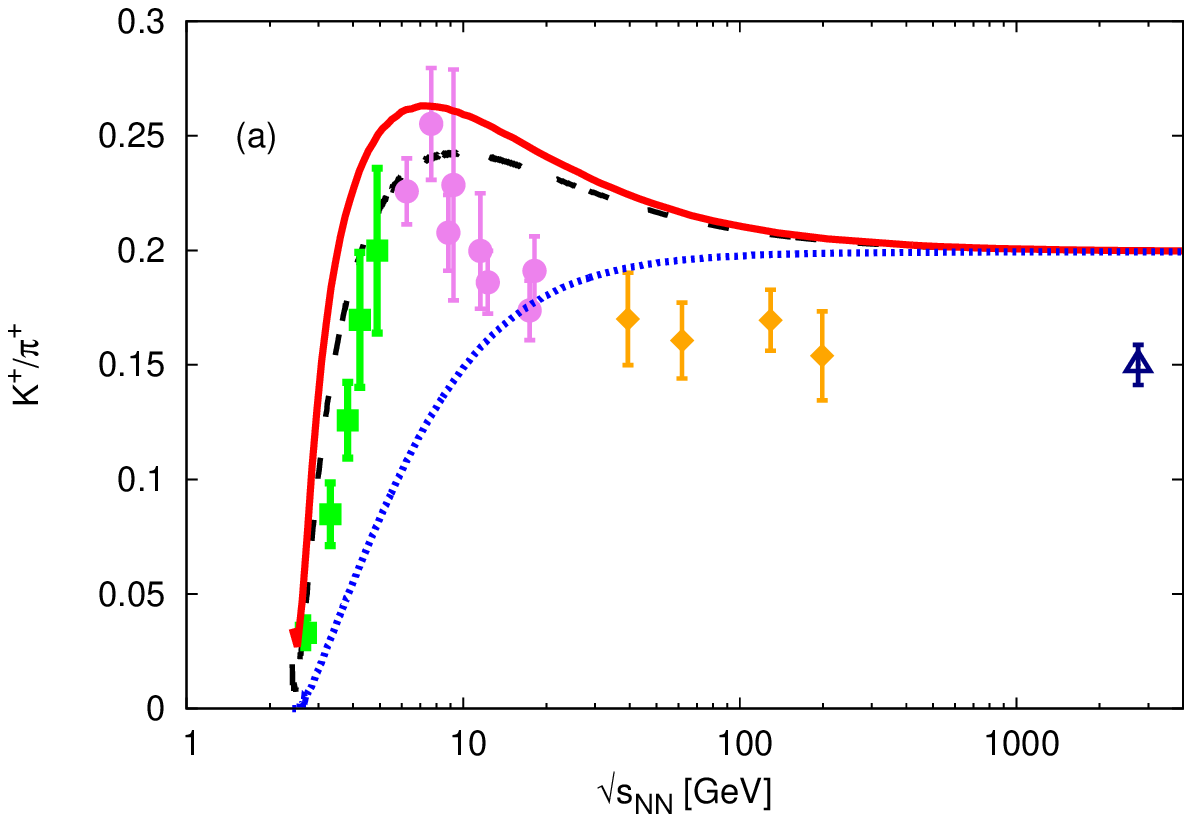}
\includegraphics[width=5.cm]{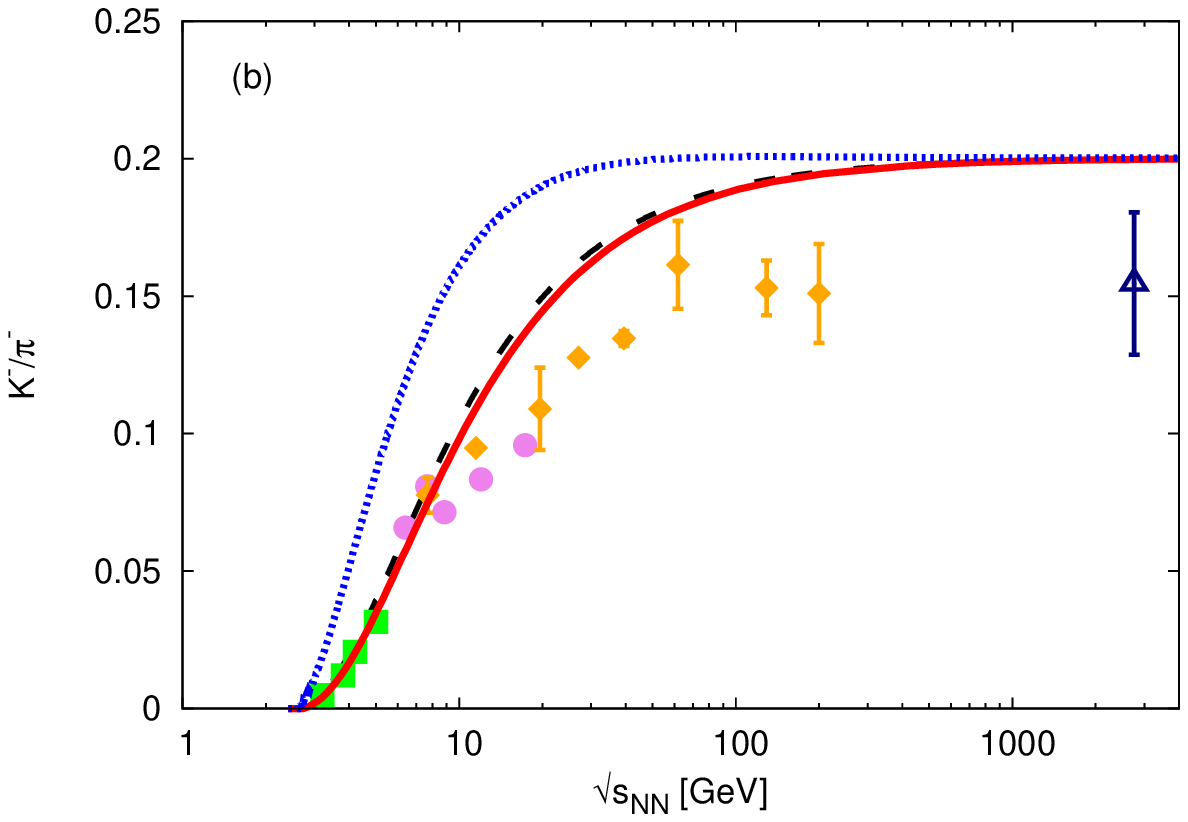} 
\includegraphics[width=5.cm,angle=-0]{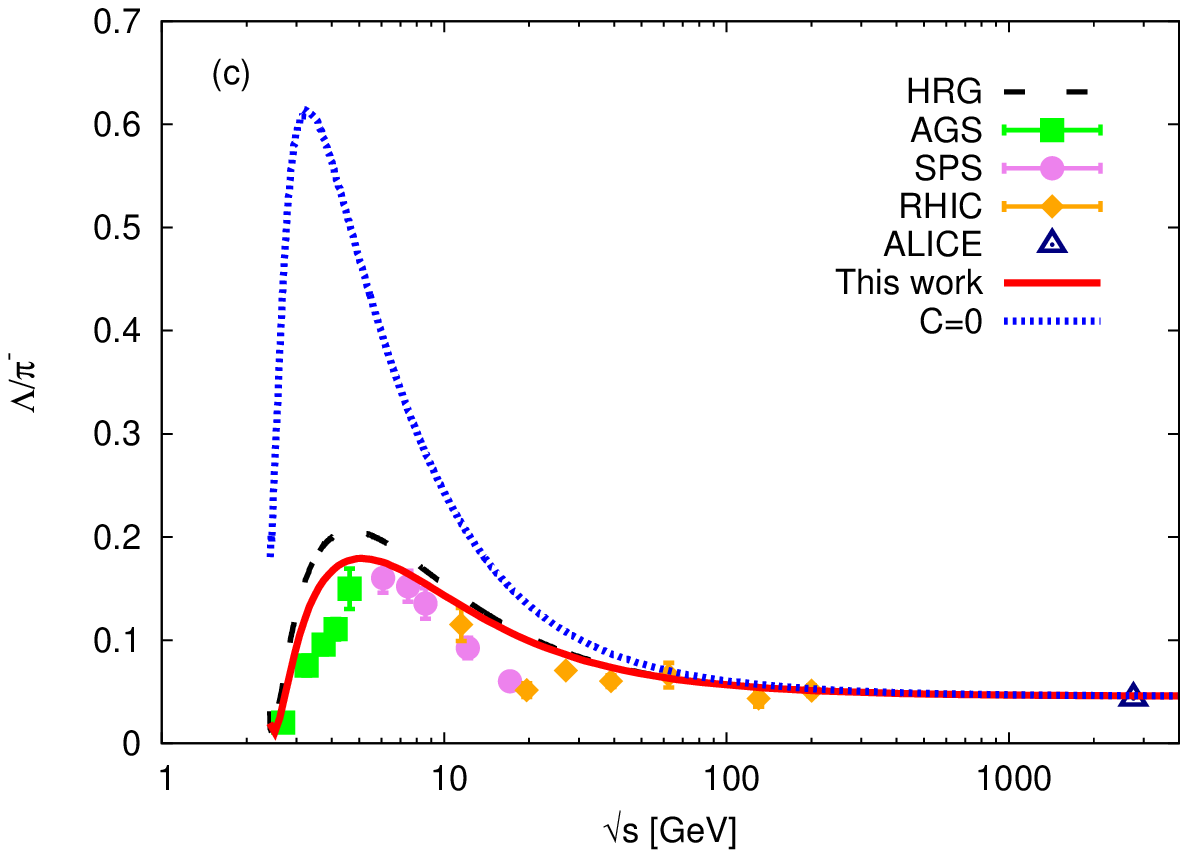} 
\includegraphics[width=5.cm,angle=-0]{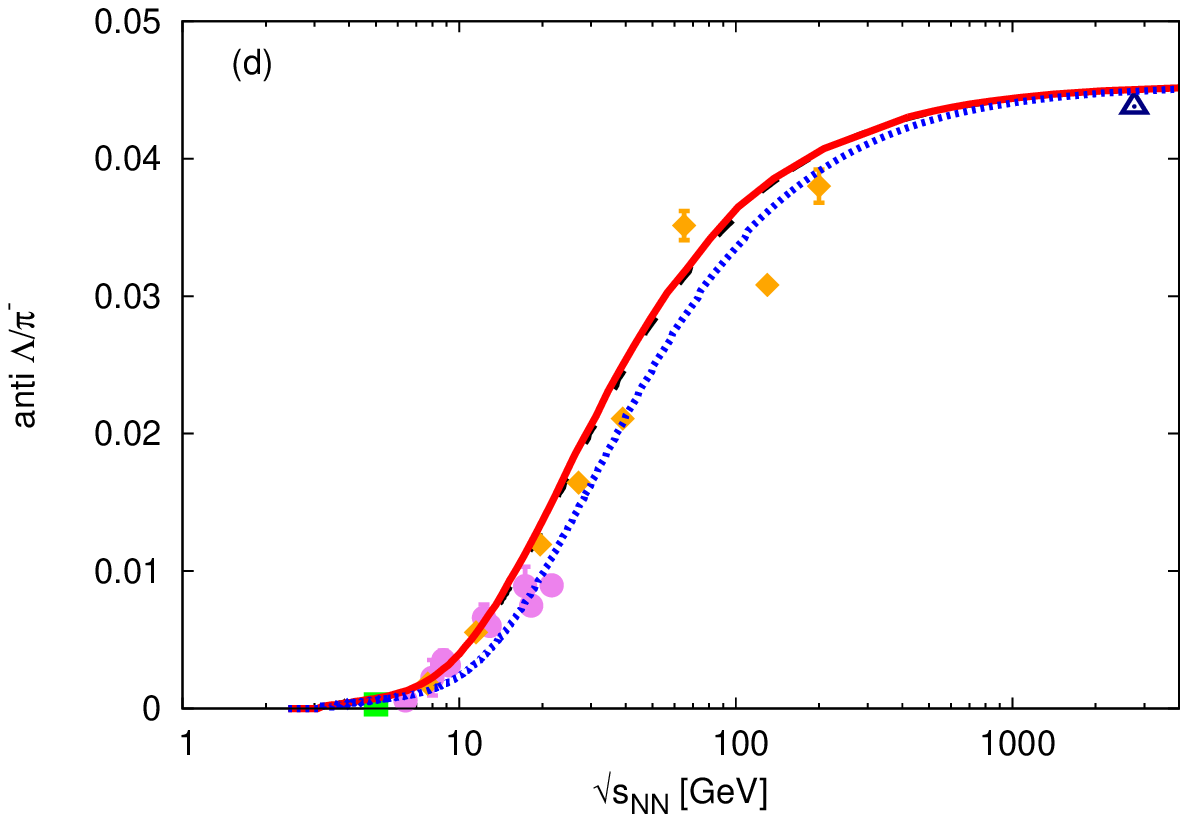} 
\includegraphics[width=5.cm,angle=-0]{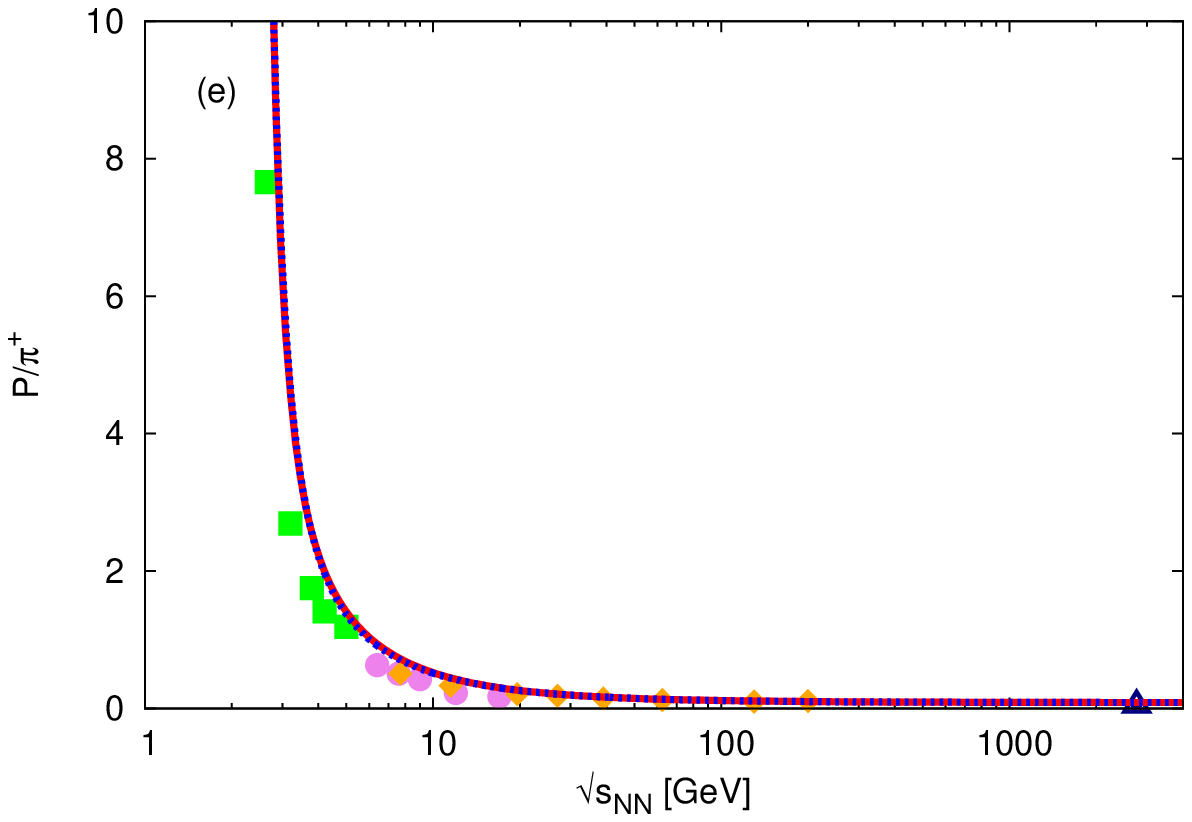}
\includegraphics[width=5.cm,angle=-0]{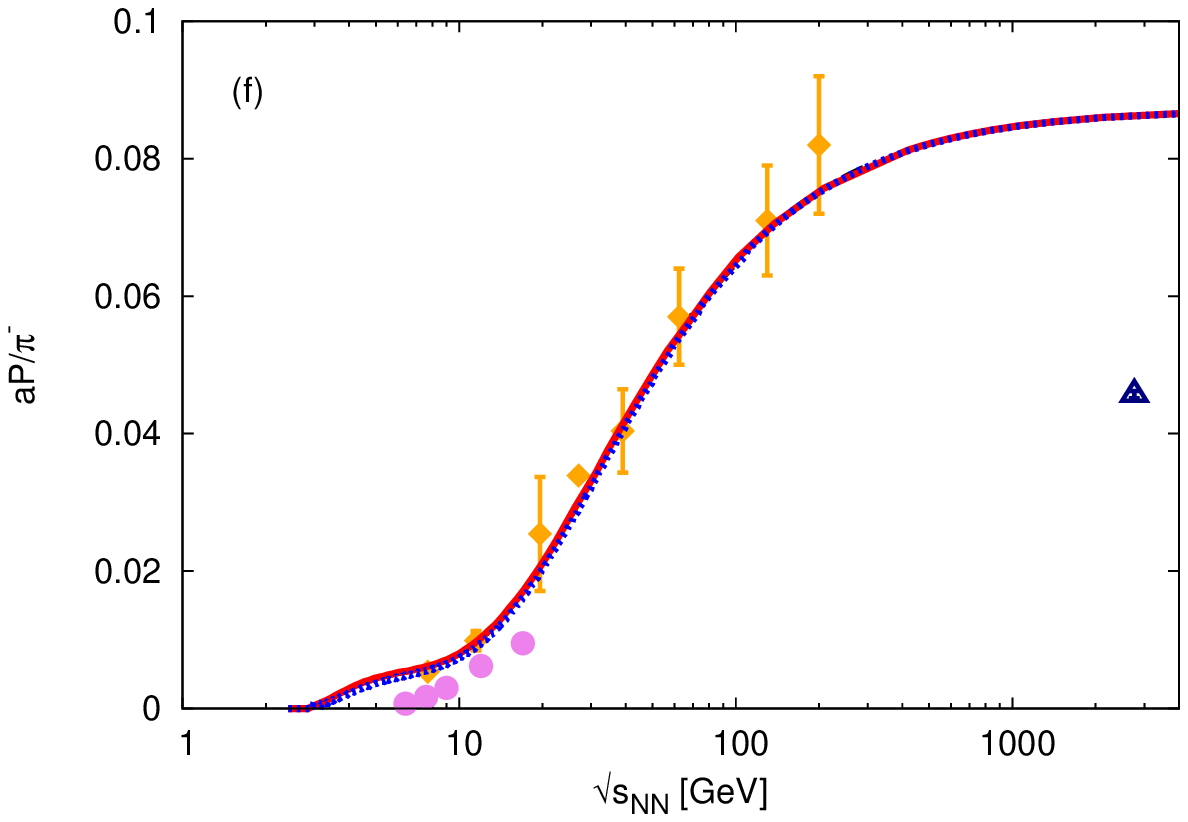} 
\includegraphics[width=5.cm,angle=-0]{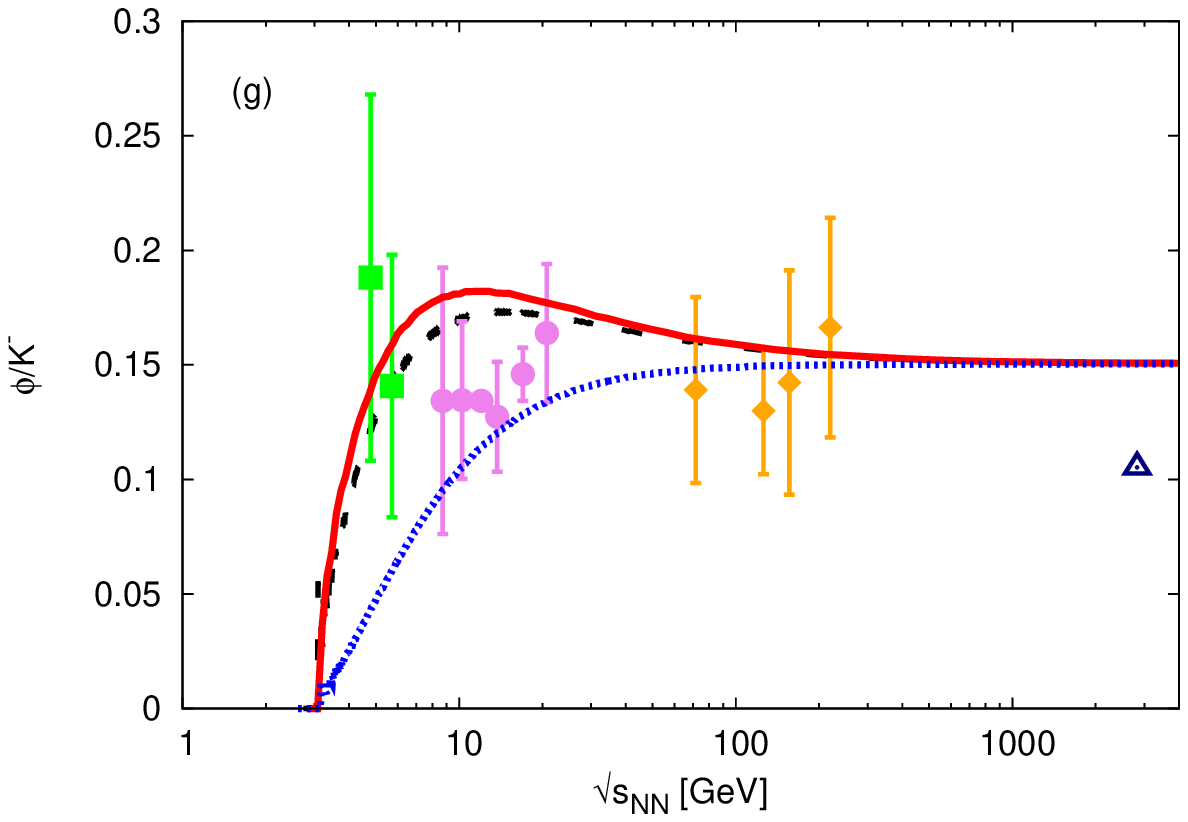} 
\includegraphics[width=5.cm,angle=-0]{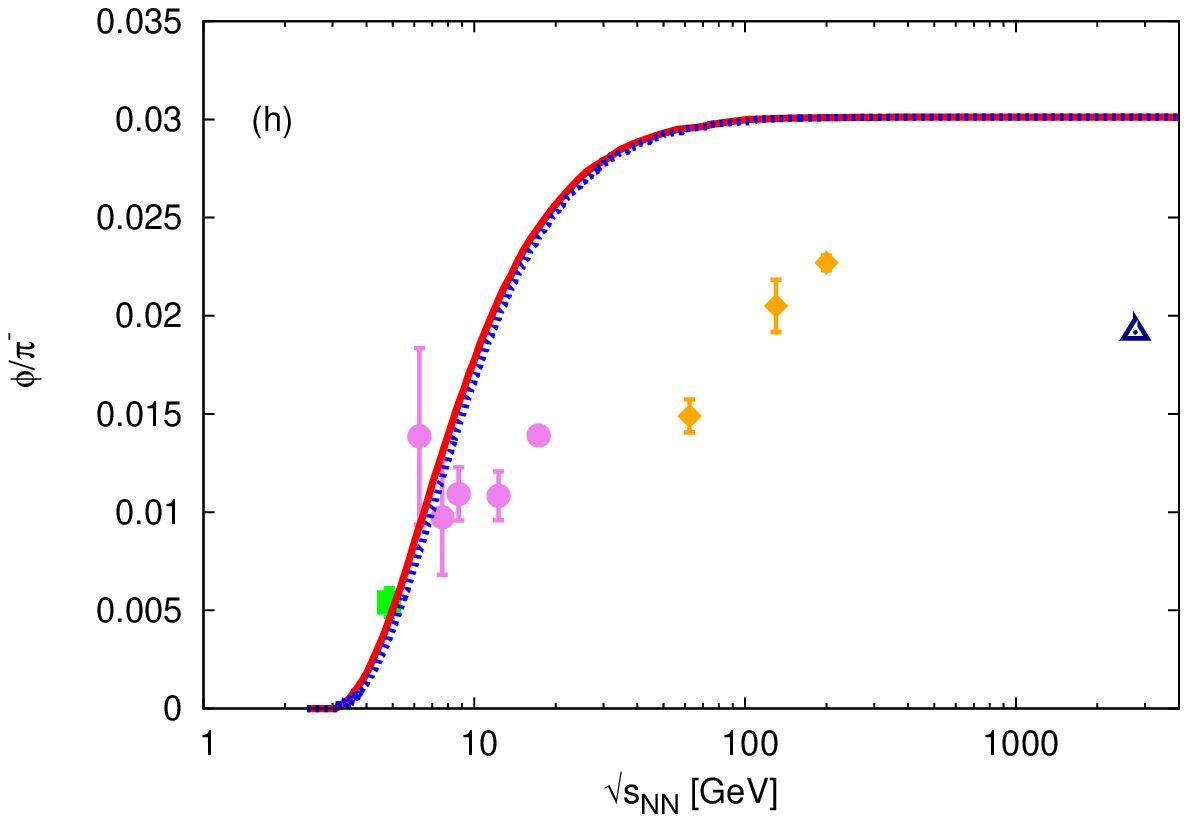} 
\includegraphics[width=5.cm,angle=-0]{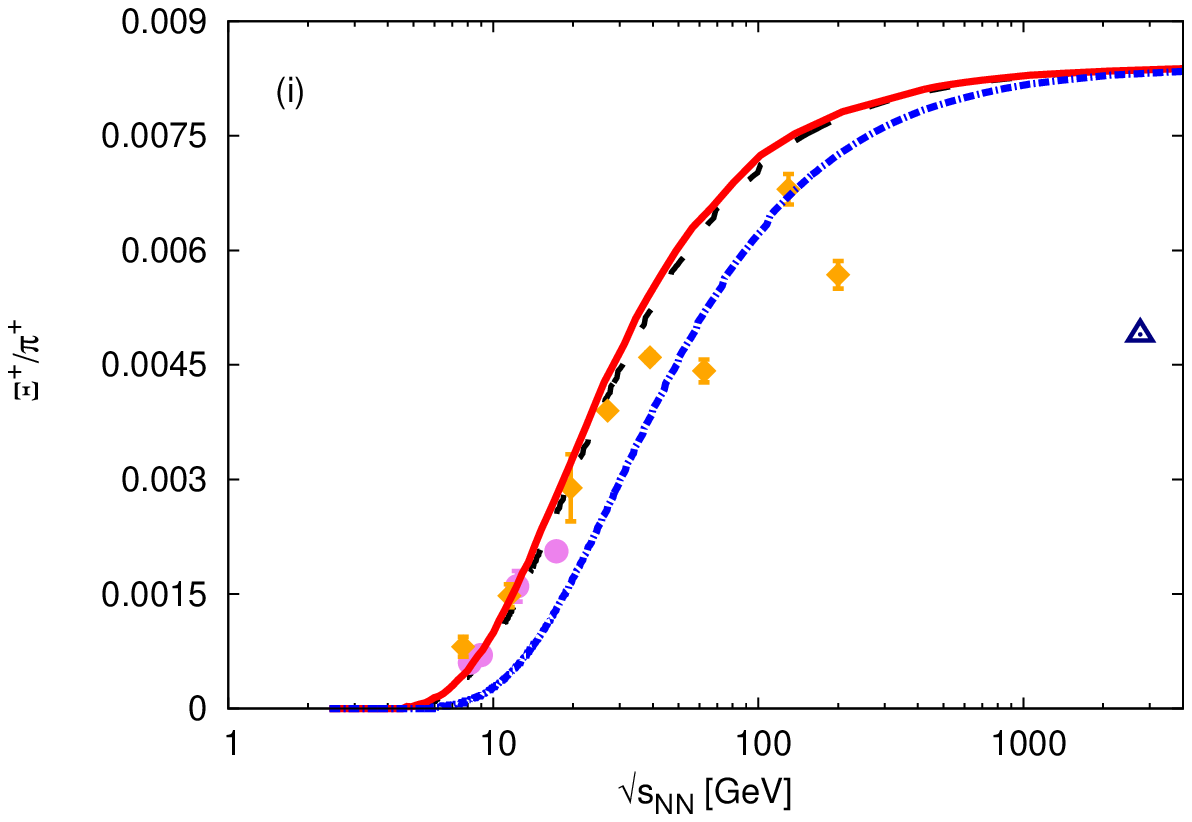}
\includegraphics[width=5.cm,angle=-0]{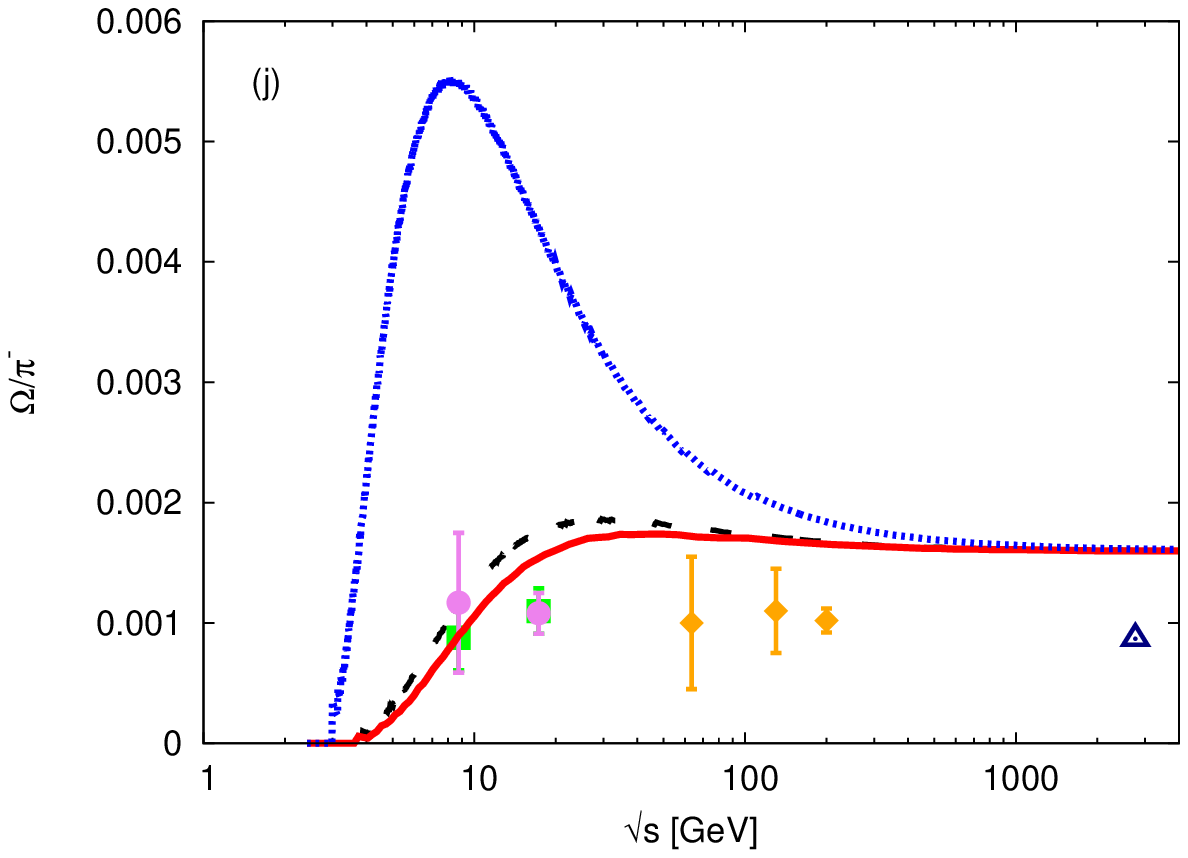}
\caption{The particle ratios $\mathrm{K}^+/\mathrm{\pi}^+$ (a), $\mathrm{K}^-/\mathrm{\pi}^-$ (b), $\mathrm{\Lambda}/\mathrm{\pi}^-$ (c), $\bar{\mathrm{\Lambda}}/\mathrm{\pi}^-$ (d), $\mathrm{p}/\mathrm{\pi}^+$ (e),  $\bar{\mathrm{p}}/\mathrm{\pi}^-$ (f), $\mathrm{\phi}/\mathrm{K}^-$ (g), $\mathrm{\phi}/\mathrm{\pi}^-$ (h), $\mathrm{\Xi}/\mathrm{\pi}^+$ (i), and $\mathrm{\Omega}/\mathrm{\pi}^-$ (j) are depicted as functions of $\sqrt{s_{\mathrm{NN}}}$. Measurements (symbols) are confronted to various calculations; HRG at standard $\mu_{\mathrm{S}}$ (dashed curve), $\mu_s=0.1991\mu_{\mathrm{b}}$ (green curve) and $\mu_{\mathrm{s}}=0.1991\mu_{\mathrm{b}}+{\cal O}$ (red curve).}
\label{fig:pr}
\end{figure}

In this section, we want to examine the ability of the proposed $\mu_s$, Eq. (\ref{eq:musmubNew2}), to reproduce different particle ratios. This includes $\mathrm{K}^+/\mathrm{\pi}^+$, $\mathrm{K}^-/\mathrm{\pi}^-$, $\mathrm{\Lambda}/\mathrm{\pi}^-$, $\bar{\mathrm{\Lambda}}/\mathrm{\pi}^-$, $\mathrm{p}/\mathrm{\pi}^+$,  $\bar{\mathrm{p}}/\mathrm{\pi}^-$, $\mathrm{\phi}/\mathrm{K}^-$, $\mathrm{\phi}/\mathrm{\pi}^-$, $\mathrm{\Xi}/\mathrm{\pi}^+$, and $\mathrm{\Omega}/\mathrm{\pi}^-$. Both $T$ and $\mu_B$ are fixed at varying $\sqrt{s_{\mathrm{NN}}}$, Eq. (\ref{T_muB}) and Eq. (\ref{mu_B}), respectively, while $\mu_s$ should be determined either from \cite{Tawfik:2004sw} or the present work. We examine three approaches.

For the sake of completeness, we highlight that c are determined at the state of chemical freezeout, which is conditioned to $s/T^3=7$, where $s$ is the entropy density \cite{Tawfik:2016jzk,Tawfik:2005qn,Tawfik:2004ss}.  The entropy density can be directly derived from Eq. (\ref{eq:lnz1}), for instance. This condition excellently described both freezeout parameters $T$ and $\mu_B$ as determined from the statistical fits of different particle ratios measured at various collision energies.

Figure \ref{fig:pr} presents different particle ratios as functions of  $\sqrt{s_{\mathrm{NN}}}$. Various HRG-calculations (curves) are confronted to the measurements (symbols). The calculations at {\it standard} $\mu_{\mathrm{S}}$ are depicted by the dashed curves. Here, $\mu_{\mathrm{S}}$ is calculated at $T$ and $\mu_{\mathrm{B}}$, which - in turn - vary with $\sqrt{s_{\mathrm{NN}}}$. The value of $\mu_{\mathrm{S}}$ assuring averaged strangeness conservation is the one that enters our present calculations - besides other parameters including $T$ and $\mu_{\mathrm{B}}$, etc. - for the given particle ratios. 

The present calculations which are performed at $\mu_{\mathrm{s}}=0.1991\mu_{\mathrm{b}}+{\cal O}$, Eq. (\ref{eq:musmubNew2}), are given by the solid curves. Here, $\mu_{\mathrm{s}}$ is first fixed to $\sim20\%\mu_{\mathrm{b}}$. Then, the additional quantity ${\cal O(T(\mu_{\mathrm{b}}))}$ should be determined in order to assure strangeness conservation. 

We notice that both curves excellently agree with each other. In other words, the proposed phenomenological approach, Eq. (\ref{eq:musmubNew2}), is well approved, at least, that it excellently reproduces the well-known results of the statistical-thermal approaches. 

We also draw results based on constant $\mu_s=0.1991\mu_b$ (dotted curve). We observe that $50\%$ of the particle ratios are well reproduced, while this entirely fails in the remaining $50\%$. This illustrates the importance of ${\cal O}$, Fig. \ref{fig:mubSqrts}(b), which becomes significant at large $\mu_{\mathrm{b}}$ or low $\sqrt{s_{\mathrm{NN}}}$. At high $\sqrt{s_{\mathrm{NN}}}$, ${\cal O}(T(\mu_{\mathrm{b}}))$ becomes negligibly small, i.e. $\mu_{\mathrm{s}}=0.1991\mu_{\mathrm{b}}$ seems to work well in this limit. This might shed lights on the strangeness production at varying collision energies.

\subsubsection{RHIC particle yields and ratios}
\label{sec:pRtYl}

\begin{figure}[!htb]
\includegraphics[width=8.cm]{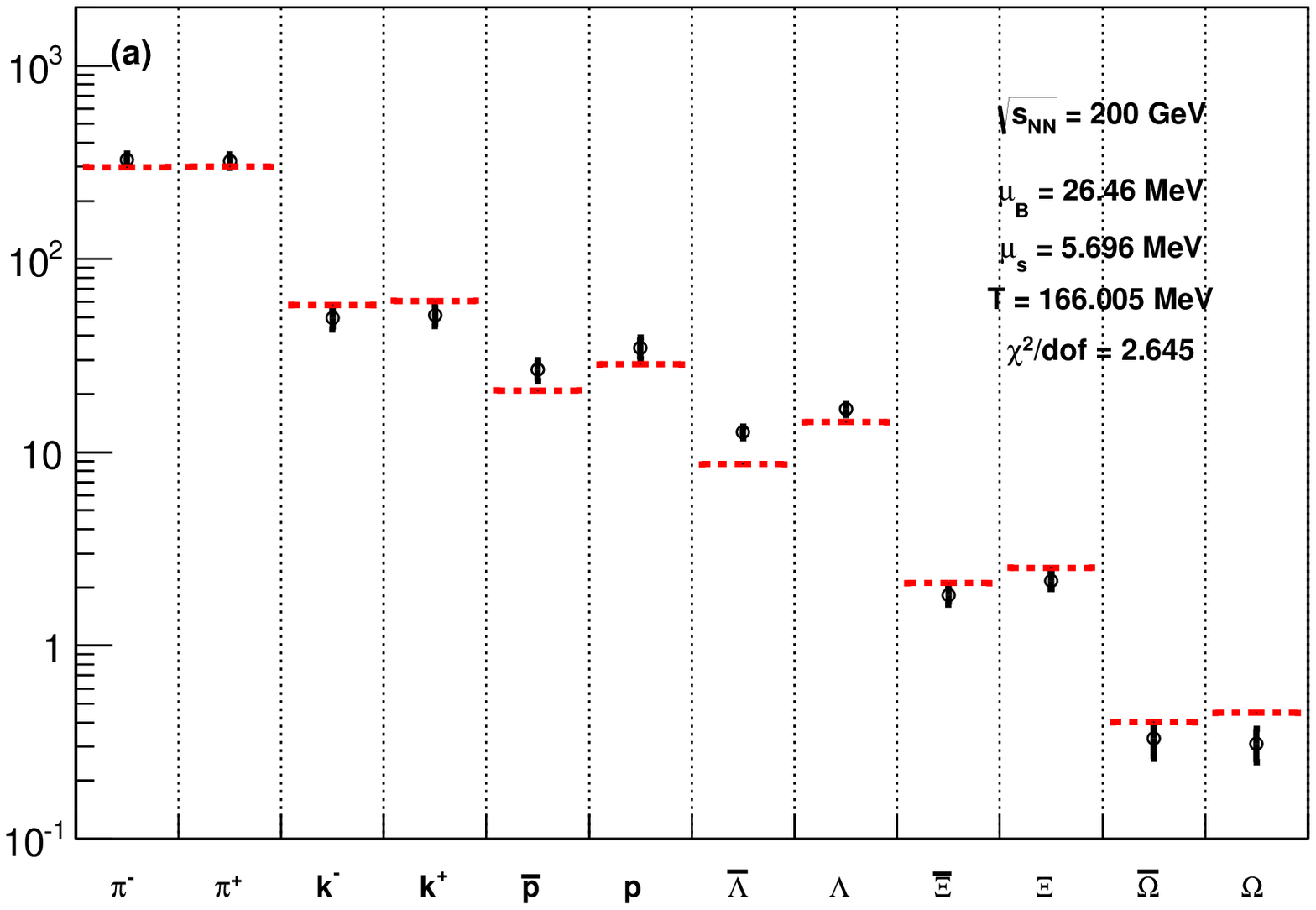}
\includegraphics[width=8.cm]{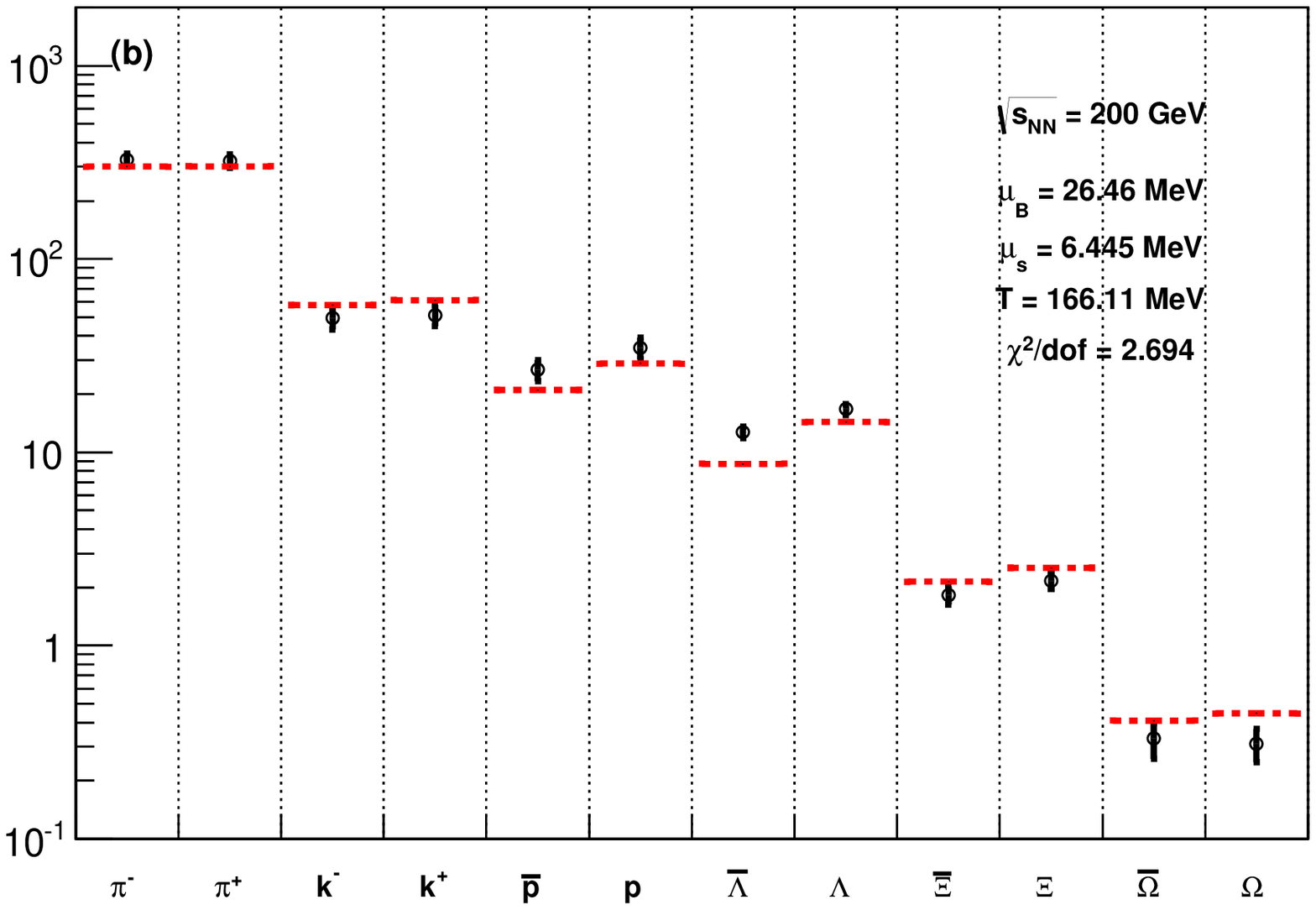}\\
\includegraphics[width=8.cm]{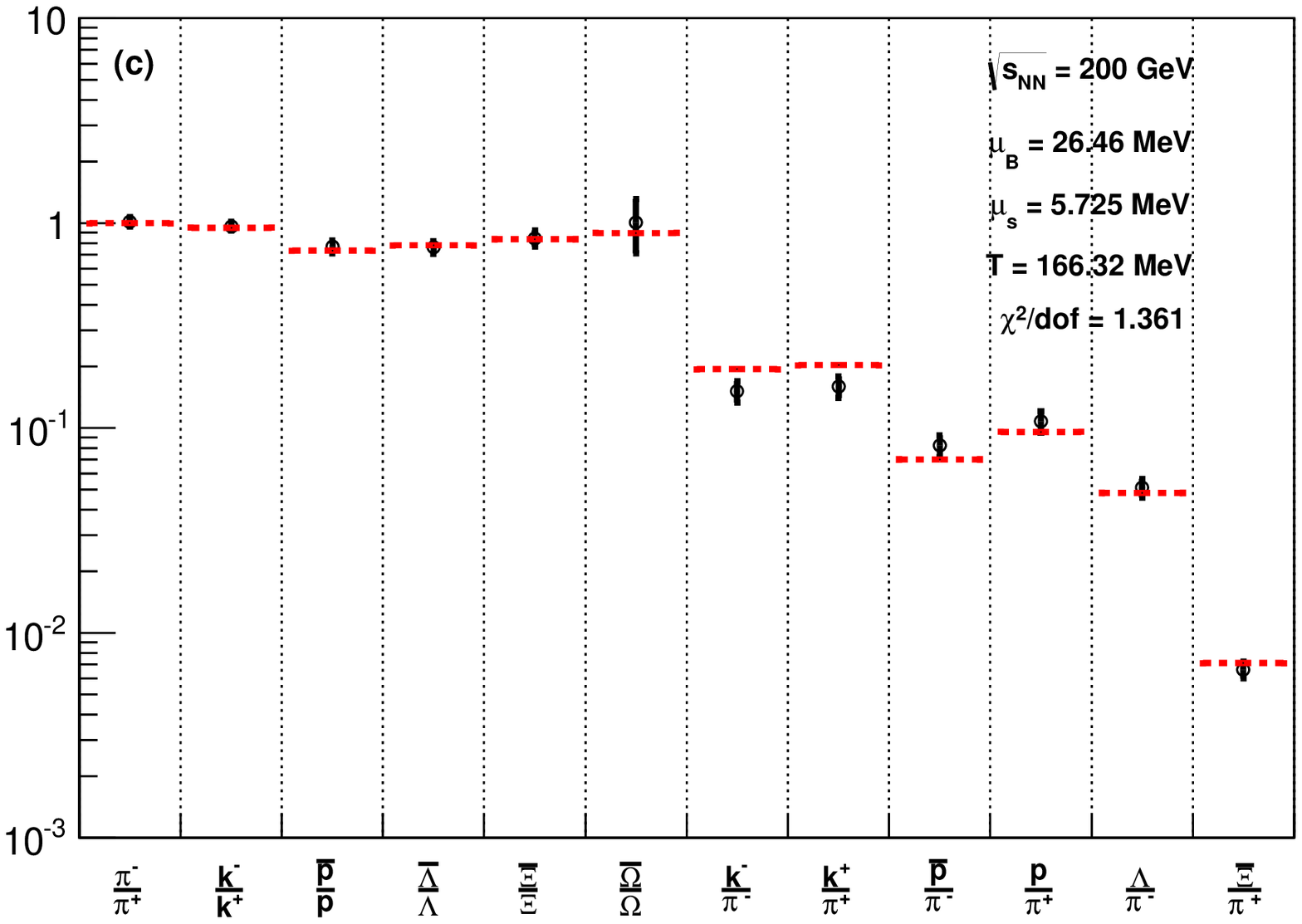}
\includegraphics[width=8.cm]{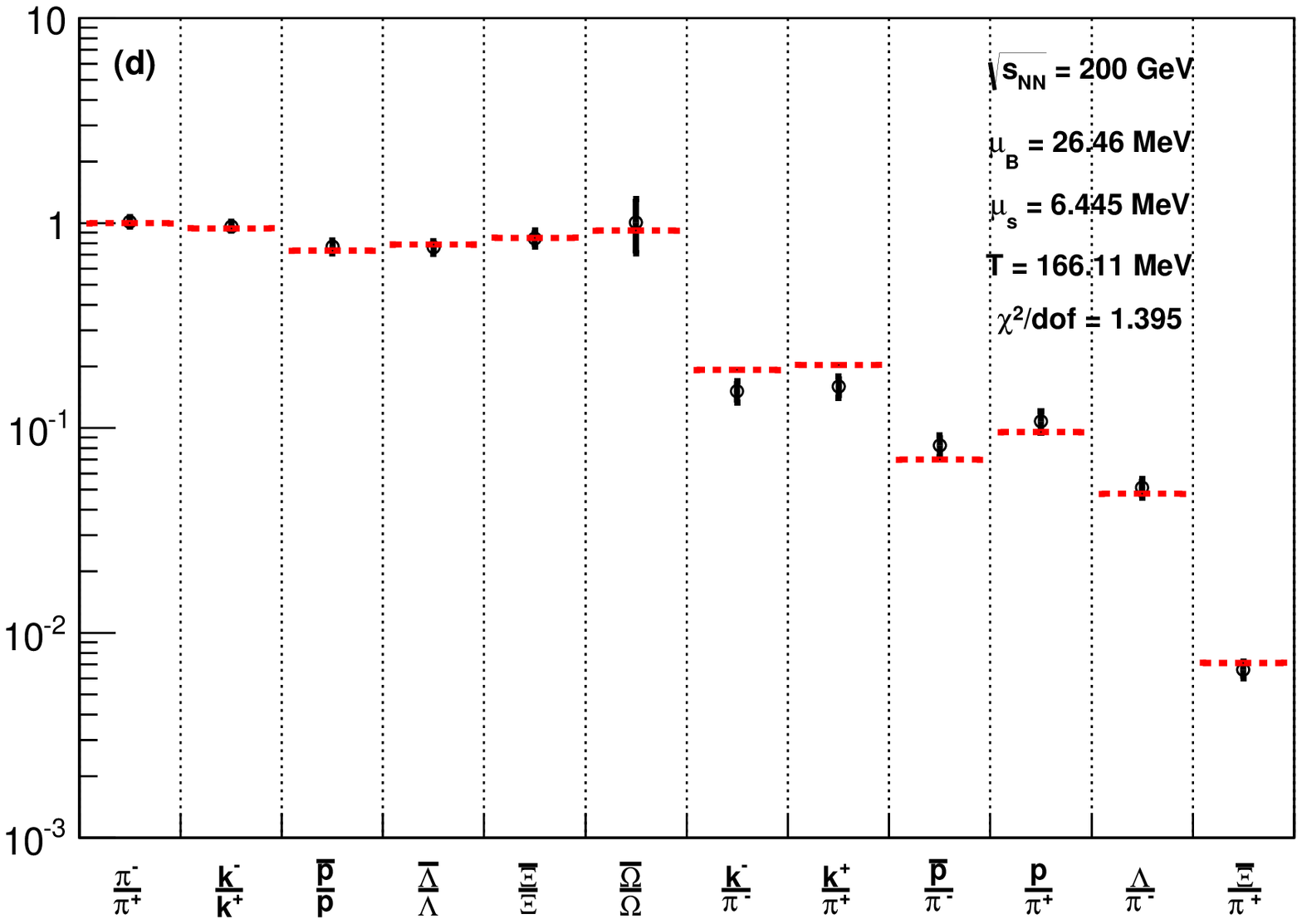} 
\caption{Particle yields (top panel) and ratios (bottom panel) measured at $200~$GeV are fitted to various calculations, where $\mu_{\mathrm{S}}$ is statistically (left-hand panel) and partly phenomenologically determined, present approach (right-hand panel). Other freezeout parameters, e.g. $T$ and $\mu_{\mathrm{B}}$, are taken as free parameters. The goodness of both fits is measured by corresponding $\chi^2/$dof.}
\label{fig:C}
\end{figure}

\begin{figure}[!htb]
\includegraphics[width=8.cm]{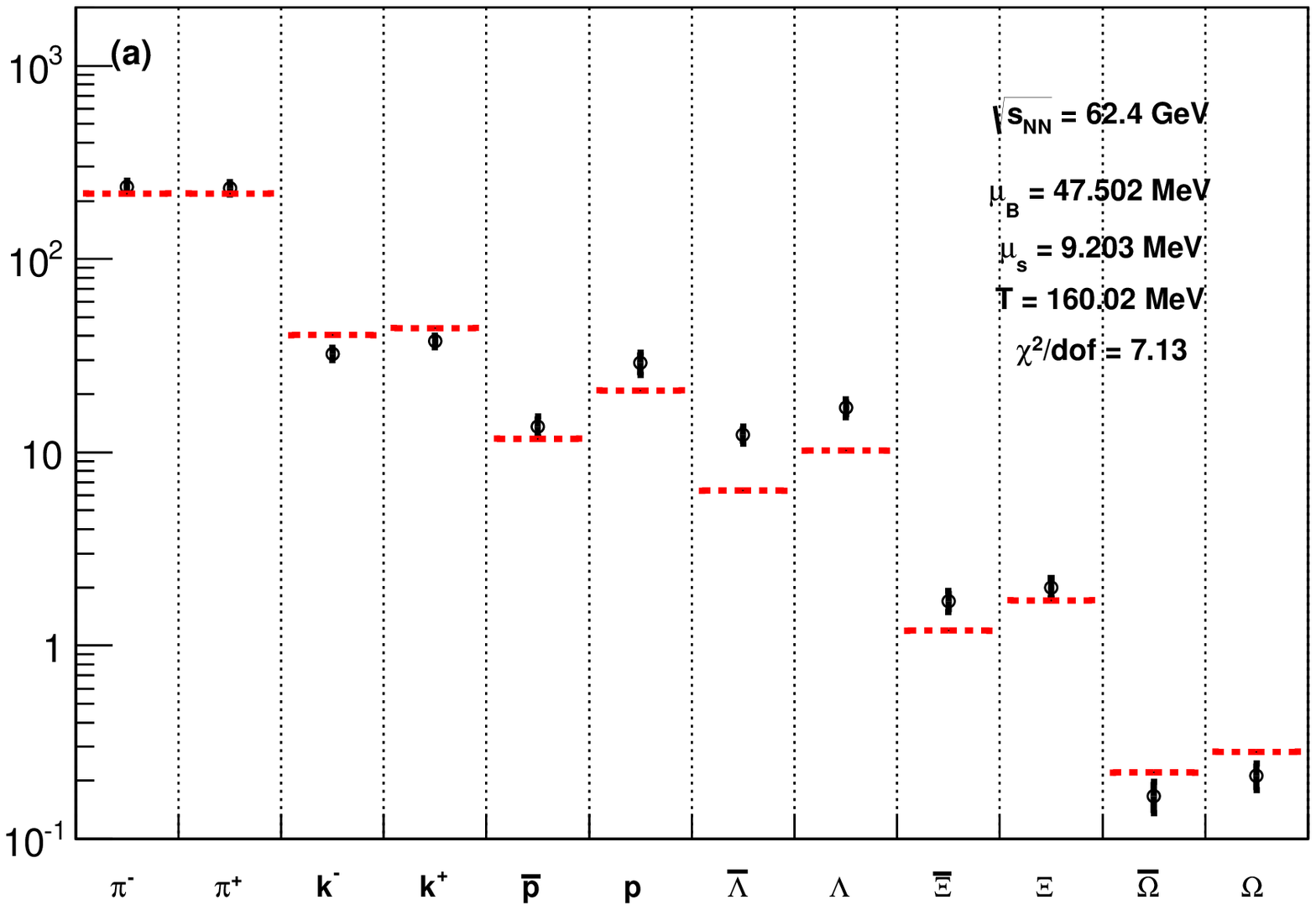}
\includegraphics[width=8.cm]{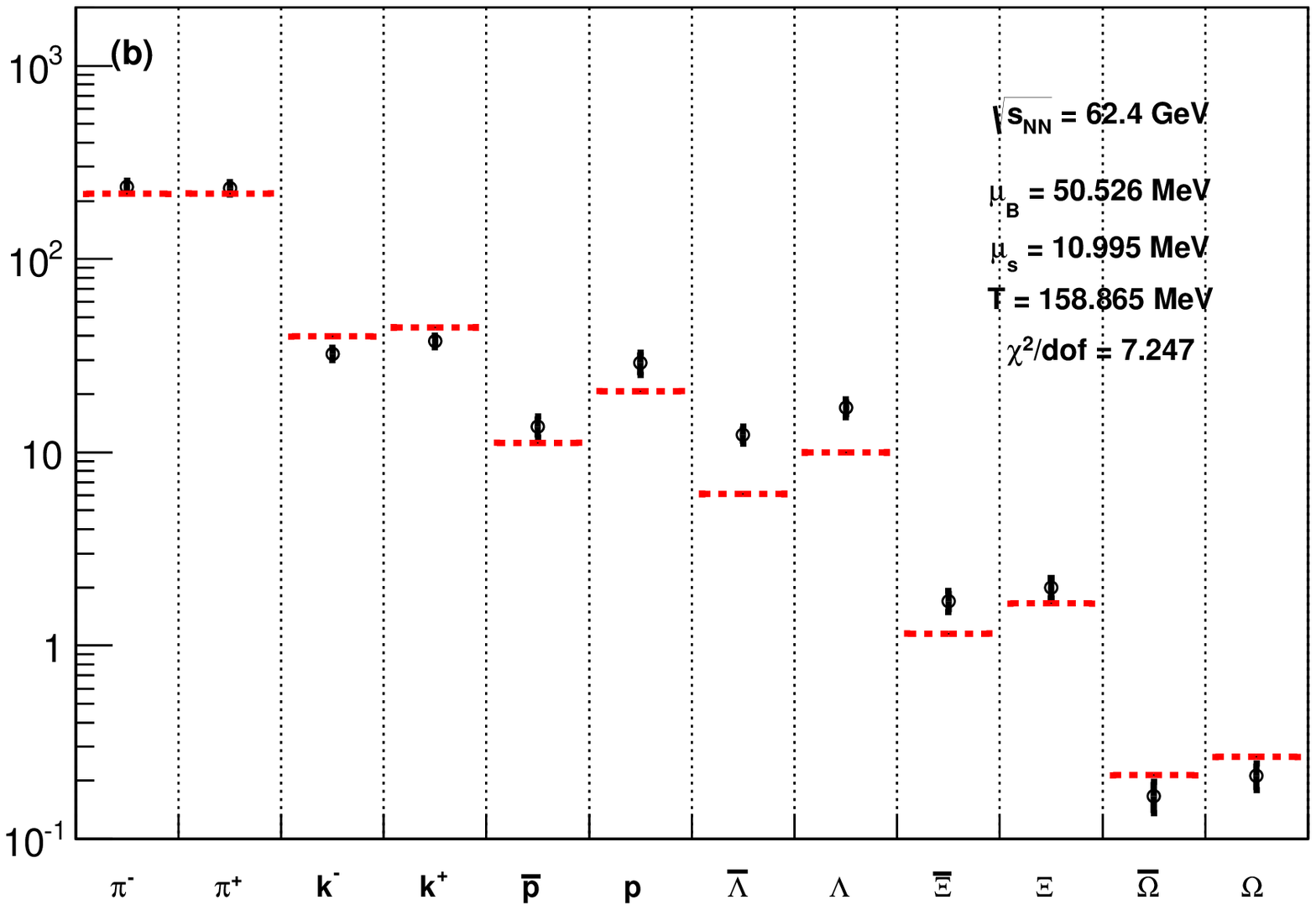}\\
\includegraphics[width=8.cm]{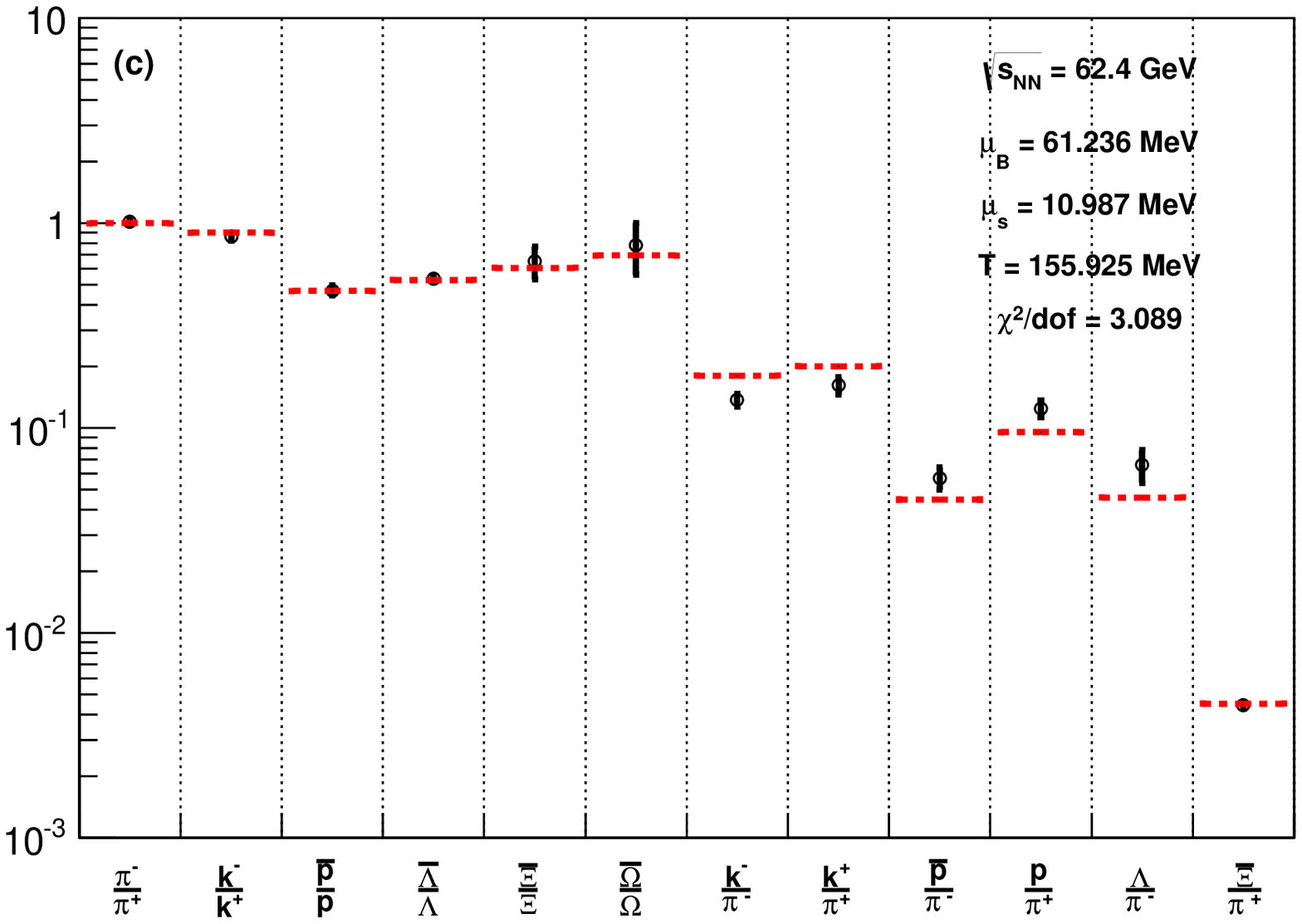}
\includegraphics[width=8.cm]{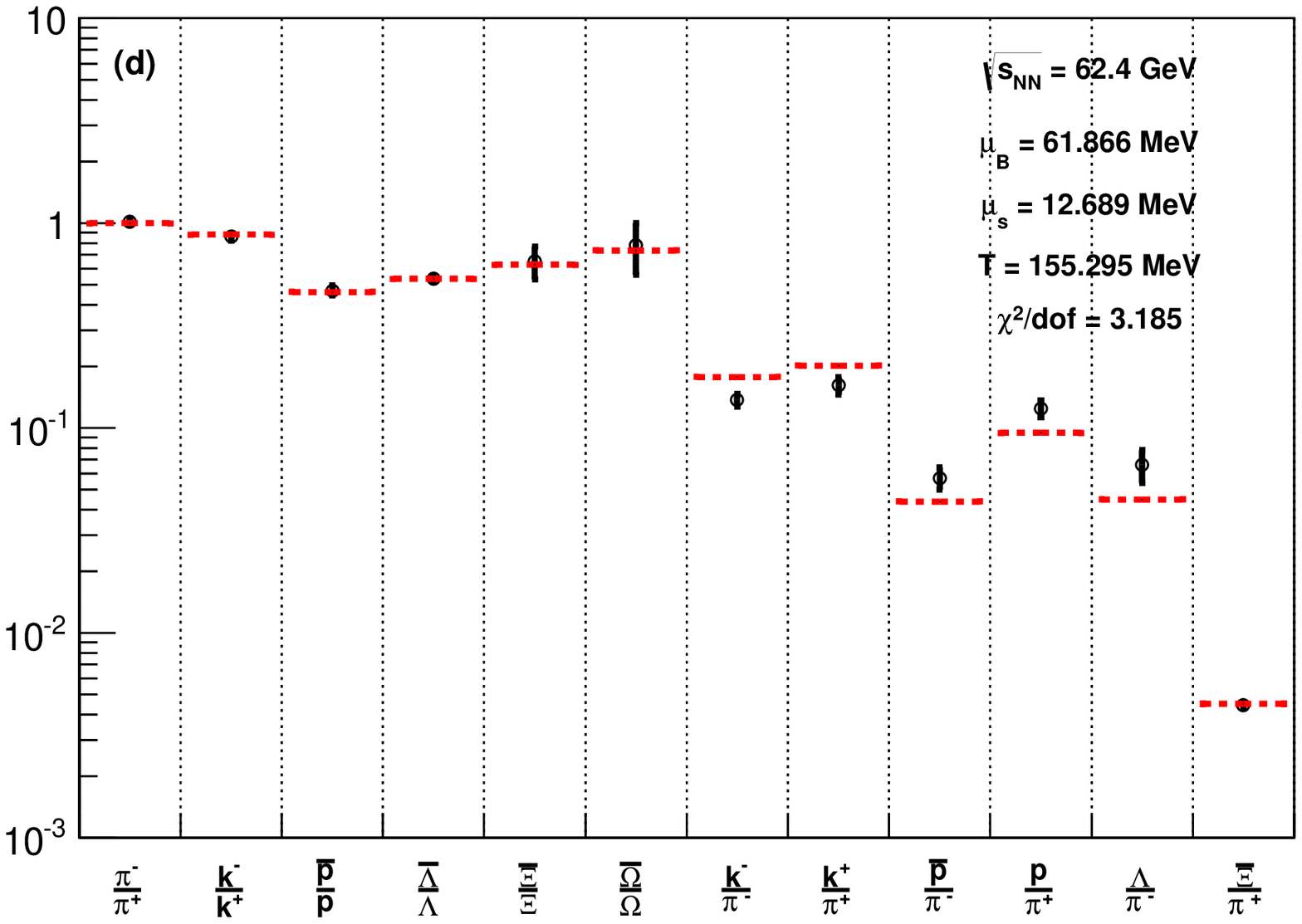} 
\caption{The same as in Fig. \ref{fig:C} but at $62.4~$GeV.}
\label{fig:C2}
\end{figure}

The next check to be made out in assuring best parameterization of ${\cal O}(T(\mu_{\mathrm{b}}))$ is a direct comparison between the present calculations, where $\mu_{\mathrm{s}}$ is partly phenomenologically  fixed to $\mu_{\mathrm{b}}$ from statistical-thermal calculations \cite{Tawfik:2014eba,Tawfik:2004sw}. Possible impacts of ${\cal O}(T(\mu_{\mathrm{b}}))$, Eq. (\ref{eq:musmubNew}), is illustrated in Figs. \ref{fig:C} and \ref{fig:C2}. Top panels depict the statistical fits of various particle yields, while the bottom ones give the results from fits of different particle ratios, at $\sqrt{s_{NN}}=200$ and $62.4~$GeV, respectively. The left- and right-hand panels distinguish between the two different procedures in determining and utilizing $\mu_{\mathrm{s}}$ (or equivalently $\mu_{\mathrm{S}}$). This aims to elaborate how ${\cal O}(T(\mu_{\mathrm{b}}))$, Eq. (\ref{eq:musmubNew}), is efficient in reproducing the particle yields and ratios. We simply compare between results performed at $\mu_{\mathrm{s}}$ that is based on the present approach, Eq. (\ref{eq:musmubNew2}), and results based on the other procedure, which is widely utilized in the statistical thermal models \cite{Tawfik:2014eba,Tawfik:2004sw}. 

We notice that both approaches excellently agree with each other. Top panels illustrate that they  almost equivalently fit the particle yields, at $\sqrt{s_{NN}}=200$ and $62.4~$GeV. The fitting parameters are given in the top left corners. It is found that the fireball volume and radius are also identical, 
\begin{itemize}
\item at $\sqrt{s_{NN}}=200~$GeV: $V_{\mathrm{fb}}=56.47~$fm$^3$ and $R_{\mathrm{fb}}=2.38~$fm, and
\item at $\sqrt{s_{NN}}=62.4~$GeV: $V_{\mathrm{fb}}=53.62~$fm$^3$ and $R_{\mathrm{fb}}=2.34~$fm.
\end{itemize}
To show some comparisons at different energies, we present results at $\sqrt{s_{NN}}=200$ and $62.4~$GeV. Some $\sqrt{s_{NN}}$-effects on $\mu_s$, $\mu_B$, and $T_{\mathrm{ch}}$ can be understood from both figures and the fit parameters shown in their top right corners.

From bottom panels, we conclude that  both approaches lead to almost identical fits of the particle ratios. Details about the resulting fit parameters are outlined in the top left corners.

Figure \ref{fig:C} implies that the resultant $\mu_{\mathrm{s}}$, which is partly phenomenologically fixed to $\mu_{\mathrm{b}}$ agrees well with $\mu_{\mathrm{s}}$ (or equivalently $\mu_{\mathrm{S}}$)  that was statistically determined in statistical-thermal models \cite{Tawfik:2014eba,Tawfik:2004sw}.

\section{Conclusions}
\label{sec:cncl}

The possible asymmetry between charged mesons and antimesons can be phenomenologically highlighted, through studying the energy-dependence of the antikaon-to-kaon ratios, in which $\mu_{\mathrm{b}}$ is apparently eliminated, while strangeness sector $\mu_{\mathrm{s}}$ becomes dominant. On the other hand, the energy-dependence of the ratio of antiproton-to-proton to antikaon-to-kaon is assumed to reveal fruitful information about the production of antibaryons and baryons, as the (anti)protons are dominant baryons. Furthermore, it is assumed that the dependence of the ratio of antiproton-to-proton ratio to antikaon-to-kaon entirely manifests the characteristics of the antihadron-to-hadron asymmetry observed in heavy-ion collisions. The present work presents an ambiguous analysis of four antibaryon-to-baryon ratios in dependence on the antikaon-to-kaon ratio measured in a wide range of collision energies. Hence, a genuine estimation for $\mu_{\mathrm{s}}$, which is found sensitive to both freezeout parameters, temperature ($T$) and baryon chemical potential ($\mu_{\mathrm{b}}$).

The main results can be summarized as follows. $\sim20\%$ of the baryon chemical potential ($\mu_{\mathrm{b}}$) is to be assigned to the strangeness chemical potential ($\mu_{\mathrm{s}}$). This parameterization assures strangeness conservation, especially at high collision energies. At lower collision energies, an additional functionality depending on the freezeout parameters should be added. As a direct implication, we found that the resultant $\mu_{\mathrm{s}}$ reproduces well various particle yields and ratios. Also when comparing results based on this parameterization, with corresponding results based on the statistical-thermal approaches, an excellent agreement is obtained.

One of the major questions to be answered is which significant role plays the proposed additional functionality, which as mentioned depends on the freezeout parameters, $T$ and $\mu_{\mathrm{b}}$, in characterizing the possible strangeness enhancement with increasing energies, as recently reported by the ALICE collaboration in $pp$ collisions \cite{Alice2017}. This phenomenon was first featured in the AGS and SPS results \cite{SE} and from quite uncommon
statistical mechanics features \cite{Sollfrank}. It was believed that the strangeness enhancement is likely pronounced in heavy-ion collisions, especially at low energies. The recent ALICE finding is therefore clear cut. Although the functionality that should be added to fifth of $\mu_{\mathrm{b}}$ vanishes at very high energies (very small $\mu_{\mathrm{b}}$), it linearly increases with the increase in the collision energy. On the other, at lower energies, (large $\mu_{\mathrm{b}}$), there is a rapid characteristic increases, so that the functionality largely exceeds the first term, $\sim0.2 \mu_b$.  

The energy-dependence of the functionality parameterizing $\mu_{\mathrm{s}}(\mu_{\mathrm{b}})$ and simultaneously assuring strangeness conservation, qualitatively looks very similar to the energy-dependence of the freezeout temperature \cite{Tawfik:2014eba}. The only different is that one of both quantities should be flipped, horizontally. Opposite to the resulting functionality, the freezeout temperature reaches maximum, as low $\mu_{\mathrm{b}}$. Also, we noticed that it slightly decreases with the increase in $\mu_{\mathrm{b}}$. At very high $\mu_{\mathrm{b}}$, there is a rapid decrease. At high $\mu_{\mathrm{b}}$, we realize from Eq. (\ref{eq:musmubNew2}) that the functionality, which is given in the second term in rhs, is proportional to $T(\mu_{\mathrm{B}})$. In dedicating the given energy-dependence of the entire term, the denominator plays an important role. It is obvious that this is the case as long as $\mu_{\mathrm{b}}<(b/a)^{1/(1+c)}$. It is apparent that his range of $\mu_{\mathrm{b}}$ covers (agrees well with) the baryon chemical potential characterizing the entire hadron phase. At higher baryon chemical potential, $\mu_{\mathrm{s}}$ clearly diverges. 

We finally conclude that $\mu_{\mathrm{s}}$ seems to have strong two-fold dependency on $\mu_{\mathrm{b}}$. First, it is one-fifth $\mu_{\mathrm{b}}$. Second, it is directly proportional to $T(\mu_{\mathrm{B}})$ normalized to $f(\mu_{\mathrm{b}})= a_1\, \mu_{\mathrm{b}}^{-b_1} - c_1\, \mu_{\mathrm{b}}$. The latter qualitatively describes the strangeness suppression (enhancement) with the increase in the baryon chemical potential (the collision energy). This observation is dictated by $T(\mu_{\mathrm{B}})$, i.e. the freezeout boundary, while the $\mu_{\mathrm{b}}$ is obviously correlated with $\mu_{\mathrm{s}}$, as the strange quarks possess both types of chemical potentials.

To summarize, we found that the proposed funtionality suggests strangeness enhancement with the increase in the collision energy. This might be capable to model the observed enhanced production of multi-strange hadrons \cite{Alice2017}. Furthermore, the proposed funtionality might replace the {\it ad hoc} assumption of statistical, e.g. canonical suppression, and/or dynamical, e.g. non-unity quark occupancy parameter, variation of the strange quarks forming hadrons. What is proposed here is the actual strangeness enhancement. A future work shall be devoted to studying whether one can distinguish between this and the canonical (dynamical) suppression due to QGP formation. For example, one would think of an answer to the question whether is any enhancement left over when one subtracts the rather mundane effect of canonical suppression?


\end{document}